\documentclass[useAMS,usenatbib]{mn2e}
\usepackage{epsfig}
\usepackage[]{natbib}
\usepackage{aas_macros}     
\usepackage{amsfonts}
\usepackage{amsmath}
\usepackage[usenames]{color}
\usepackage{url}
\usepackage[singlelinecheck=on]{caption}
\usepackage{hyperref}

\definecolor{darkgreen}{cmyk}{0.85,0.2,1.00,0.2}

\newcommand{\rL}{r_{\rm L}}
\newcommand{\rNL}{r}
\newcommand{\Vdn}{$V {\rm d}n$}
\newcommand{\rhov}{\rho_{\rm v}}
\newcommand{\rhom}{\rho_{\rm m}}
\newcommand{\nv}{n_{\rm v}}
\newcommand{\nh}{n_{\rm h}}

\newcommand{\deltac}{\delta_{\rm c}}
\newcommand{\deltav}{\delta_{\rm v}}

\def\be{\begin{equation}}
\def\ee{\end{equation}}
\def\ba{\begin{eqnarray}}
\def\ea{\end{eqnarray}}

\title[The abundance of voids and the excursion set formalism]
{The abundance of voids and the excursion set formalism}
\author[Elise~Jennings, Yin~Li, Wayne Hu]
{Elise~Jennings$^{1}$\thanks{E-mail: ejennings@kicp.uchicago.edu}, Yin Li$^{1,2}$, Wayne Hu$^{1,3}$ \\
$^{1}$ The Kavli Institute for Cosmological Physics and Enrico Fermi Institute, University of Chicago, 5640 South Ellis Avenue, \\ Chicago, IL 60637, U. S.\\
$^{2}$ Department of Physics, University of Chicago, 5720 South Ellis Avenue, Chicago, IL 60637, U. S.\\
$^{3}$ Department of Astronomy and Astrophysics, University of Chicago, 5640 South Ellis Avenue, Chicago, IL 60637, U. S.
}

\begin{document}

\date{\today}

\pagerange{\pageref{firstpage}--\pageref{lastpage}} \pubyear{2013}

\maketitle

\label{firstpage}

\begin{abstract}
We present measurements of the number density of voids in the dark matter distribution from a series of N-body simulations
of a $\Lambda$CDM cosmology. We define voids as spherical regions of $\rhov = 0.2\rhom$ around density minima in order 
to relate our results to the predicted abundances using the excursion set formalism.
Using a linear underdensity of $\deltav = -2.7$, from a spherical evolution model, we find that a volume conserving model, which does not conserve number density in the mapping from the linear to nonlinear regime,  matches the measured abundance to within 16\% for a range of void radii $1< r(h^{-1}$Mpc$)<15$.
This model fixes the volume fraction of the universe which is in voids and assumes that voids of a similar size 
merge as they expand by a factor of 1.7 to achieve a nonlinear density of $\rhov = 0.2\rhom$ today.
We find that the model of \citet{2004MNRAS.350..517S} for the number density of voids greatly overpredicts the abundances over the same range of scales.
We find that the volume conserving model works well at matching the number density of voids measured from the simulations at higher redshifts, $z=0.5$ and 1, as 
well as correctly  predicting the abundances to within 25\% in a simulation of a matter dominated $\Omega_{\rm m} = 1$ universe.
We examine the abundance of voids in the halo distribution and find fewer small, $r<10 h^{-1}$Mpc, voids and many more large, $r>10 h^{-1}$Mpc,
voids compared to the dark matter. These results indicate that voids identified in the halo or galaxy distribution are related to the underlying void
distribution in the dark matter in a complicated way which merits further study if voids are to be used as a precision probe of cosmology.
\end{abstract}


\section{Introduction}

\label{sect:intro}
Galaxy redshift surveys allow us to study and map out 
the large scale structure of our Universe 
revealing a hierarchical mass distribution with substructure over a wide range of scales. 
The main components of the galaxy distribution are arranged in a remarkable 
\lq cosmic web\rq \, \citep{1996Natur.380..603B} made up of clusters of galaxies connected by filaments with large empty voids
which occupy most of the volume.
Only recently have systematic studies using voids as precision probes of the growth of structure been possible due to the increased depth
and volume of current galaxy surveys \citep{2001MNRAS.328.1039C, 2000AJ....120.1579Y, 2009ApJS..182..543A}.
In this paper we study the distribution of underdense void regions in the dark matter and halo distributions using
 N-body simulations. We focus on the excursion set method which gives an 
analytical prescription for the number density of voids which we compare with
measurements from simulations.

Voids are a common feature in galaxy surveys
with one of the most well known discoveries being the  void in Bo\"{o}tes
which has a diameter of approximately 50$h^{-1}$Mpc \citep{1981ApJ...248L..57K}. 
Since then several surveys such as the Center for Astrophysics Redshift Survey \citep{1989Sci...246..897G},
 the Southern Sky
Redshift Survey \citep{1992ApJ...390...17M} and the deeper Las Campanas Redshift
Survey \citep{1996ApJ...470..172S}
have identified voids in the distribution of galaxies and clusters confirming that they are the dominant and volume filling
component of our Universe. Most recently \citet{2012MNRAS.421..926P} and \citet{2012ApJ...761...44S} both 
used the Sloan Digital Sky
Survey Data Release 7 (SDSS DR7) \citep{2009ApJS..182..543A} to identify voids. \citet{2012MNRAS.421..926P} found 
1054 statistically significant voids with radii $r>10h^{-1}$Mpc
with an absolute magnitude cut of $M_r<-20.09$.
They argue that voids of effective radius $r_{\rm eff} \sim 20h^{-1}$Mpc dominate the void volume with the largest void
in the sample having $r\sim 30h^{-1}$Mpc.  \citet{2012ApJ...761...44S} constructed the first public void
catalog using the full extent of the SDSS DR7 spectroscopic survey which included the LRGs and found large voids in the sample of
$r\sim 50-60h^{-1}$Mpc in radius.

Early numerical and theoretical work on the evolution of voids
by \citet{1991ApJ...377...14R}, \citet{1992ApJ...388..234B} and \citet{1993ApJ...410..458D} 
focused on the expansion of initial linear undensities up to the moment of shell crossing, which is 
used to define a characteristic time in the formation of voids.
They considered spherical voids in a $\Omega_{\rm m}=1$ universe and found that shell crossing occurs at a linear underdensity of $-2.7$
at which point the comoving size of the void has increased by a factor of 1.7 \citep[see also][]{1993MNRAS.263..481V, 2001ApJ...548....1F}.
Many studies since then have focused on analysing the dynamics and statistical properties 
of voids such as the void probability function (VPF),
  the probability that a randomly placed sphere will contain no
objects \citep{1979MNRAS.186..145W}, the void filling factor, the void number density and void density 
profiles \citep[see e.g ][]{1991PhDT........84V,2002MNRAS.337.1193M,2003MNRAS.340..160B, 
2005MNRAS.360..216C,2006MNRAS.367.1629S,2009MNRAS.400.1835B,2011A&A...534A.128E, 2011IJMPS...1...41V,2012AJ....144...16K,2013MNRAS.428.3409A}.

Recent studies have looked at stacking voids in order to increase the statistical significance of weak lensing signals
\citep{2012arXiv1211.5966H,2013ApJ...762L..20K}\citep[see also][]{1999MNRAS.309..465A}, 
the Integrated Sach-Wolfe effect \citep{2013arXiv1301.5849I,2013arXiv1301.6136C} 
or to extract cosmological parameters by modelling the distortions in redshift space
\citep{2012ApJ...754..109L, 2012ApJ...761..187S, 2012MNRAS.426..440B} or as a test of modified gravity \citep{2013MNRAS.tmp..900C}.
The precision of these tests relies on many factors, for example, given a survey or numerical simulation of a certain size, how robustly 
can we measure statistics for a void of a given size; how accurately can we predict the number density of voids in the galaxy/dark matter halo 
distribution and how well do voids in the galaxy/halo distribution trace voids in the dark matter.
In this paper we address these three issues. In our discussion of a robust void finder we do not compare with all the
 other algorithms
which have been used in previous studies - our choice of void finder is 
motivated by the excursion set formalism for the abundance of voids which
we aim to test. 

In analysing both galaxy surveys and numerical simulations a wide variety of void finding algorithms have
been used to define underdense regions as voids.
\citet{2008MNRAS.387..933C} carried out the first systematic review of 13 different void finders,
identifying only two areas of agreement amongst the different algorithms;
that voids are very underdense $(\rho\sim0.05\rho_{\rm m})$ at their centres and that voids have very steep edges.
The void finding methods include the construction of proto-voids around local minima in the smoothed density field,
after separating the galaxy sample into \lq wall\rq \,
and \lq void\rq \, galaxies \citep[see e.g.][]{1997ApJ...491..421E,2002ApJ...566..641H}; merging proto-voids which results in non-spherical voids
\citep{2005MNRAS.360..216C}; the watershed algorithm \citep{2007MNRAS.380..551P} which uses the DTFE method
\citep{2007PhDT.......433S, 2011ascl.soft05003C}.
 The watershed void finder identifies minima in the density field and construct voids by flooding basins until the
\lq landscape\rq \, resembles a segmented plane where the edges of each segment outline a void region.
A similar algorithm, which we make use of in this paper, is the {\sc ZOBOV} \citep{2008MNRAS.386.2101N}
void finder which uses the Voronoi tessellation field
method \citep[see e.g.][]{2007arXiv0707.2877V}
to partition particles into zones, which are then joined together near density
minima, into voids.

There have been many studies of the excursion
set method to predict the
abundance of dark matter halos in the Universe \citep[see e.g.][for a review]{2007IJMPD..16..763Z}.
In comparison fewer studies have focused on testing the excursion set predictions for underdensities
in the dark matter distribution and we briefly outline some of these works here.
In applying this method to voids, \cite{2004MNRAS.350..517S} presented a model for the abundance of voids in the dark matter
 which included the influence of the larger scale environment on the formation of a void.
Their model takes into account two effects, firstly, a void of a given size may be embedded in another underdense region
which is on a larger scale, the \lq void-in-void\rq \, scenario, and secondly, a void of a
given size  could be embedded in an overdense region on a larger scale, the \lq void-in-cloud\rq \,
scenario.
\citet{2006MNRAS.366..467F} analysed how the barrier for shell crossing of a void in the galaxy distribution would differ from
the linear theory barrier for dark matter, finding that voids selected from catalogs of luminous
galaxies should be larger than those selected from faint galaxies \citep[see also][]{2007MNRAS.382..860D}.
\citet{2011PhRvD..83b3521D} consider using voids as a probe of primordial non-Gaussianity  and calculate the abundance of voids using the excursion set formalism
and the two barrier prescription of \cite{2004MNRAS.350..517S}.
\citet{2006MNRAS.367.1629S} define voids as isolated regions of the low-density excursion set
specified by density thresholds and measured the abundance and morphology of voids using N-body simulations.
In this paper we wish to test the predictions of the \cite{2004MNRAS.350..517S} excursion set
model by comparing them to
measurements of  void abundances from N-body simulations. Our definition of a void is similar to
that adopted by \citet{2006MNRAS.367.1629S} as we use a strict density threshold to define the void edge (although we
do not use isodensity contours).  To our knowledge, this is the first time that this model has been directly
compared with numerical simulations.

This paper is organised as follows.  In Section \ref{sect:esf} 
we discuss the excursion set method as it applies to dark matter halos and voids.  Appendices
\ref{sect:sphere} and \ref{sect:mod} review the salient features of the spherical evolution model that 
connects the two.
 In Section \ref{sect:void_finder} we detail our void finder and 
 the N-body simulations that were carried out. 
In Section \ref{sect:results} we present the main results of this paper on 
the number density of voids in three different cosmological models at $z=0$ and we show how this abundance changes with redshift. 
We also present the measured number density of voids in the halo distribution.
In Section \ref{sect:conclusions} we present our conclusions.

\section{Excursion set formalism for void abundance}
\label{sect:esf}

In this section we begin with a brief review of the excursion set formalism in Section \ref{sect:excursion}.
It is well known that in combination with spherical collapse this approach provides insight into many aspects of halo formation and can be used to predict dark matter halo abundances and clustering
\citep[see e.g][for a review of the subject]{2007IJMPD..16..763Z}.
The analogous spherical expansion model can likewise
be used to make excursion set predictions for voids  \citep{2004MNRAS.350..517S}. 
We review this extension in Section \ref{sect:SVdW} and show that it requires modifications
on physical grounds.   We propose a simple modification based
on volume fraction conservation in Section \ref{sect:Vdn}.

\subsection{Excursion set formalism}
\label{sect:excursion}
The excursion set formalism at its heart relies on knowledge of the statistical
properties of the linear density field.
In Fourier space, the linear density fluctuation field smoothed on a scale
$R$ is given by
\ba
\delta(\vec{x},R) = \int \frac{{\rm d}^3k} {(2\pi)^3} \delta(\vec{k}) W(\vec{k},R) e^{-i\vec{k} \cdot \vec{x}} \,,
\ea
where $\delta(\vec{k})$ is the Fourier transform of the density perturbation $\delta(\vec{x}) =  [\rho(\vec{x}) - \rhom]/\rhom$, $\rho(\vec{x})$ is the 
local density at comoving position $\vec{x}$, $\rhom$ is the background matter density and $W(\vec{k},R)$ is a filter function in Fourier space.
It is common to relate the smoothing scale $R$ to the corresponding variance of the 
linear density field
\ba
\sigma^2(R) \equiv S(R)=  \int \frac{{\rm d}k}{k} \, \frac{k^3 P(k)}{2\pi^2}  |W(k,R)|^2 \,,
\ea
where $P(k)$ is the matter power spectrum in linear perturbation theory. We can refer to a trajectory $\delta(\vec{x},S)$ as a sequence of overdensities given by subsequent increases in the smoothing scale by increments $\Delta S$. 
When a tophat filter in $k$-space is used then $\delta(\vec{x},S)$ executes a random walk. 
Given an underlying Gaussian distribution for the linear density field, the excursion set formalism 
allows us to associate  
probabilities to random walks that satisfy a given set of criteria for the smoothing scale
at which they cross various density thresholds.   Its use in defining
the statistics of objects in the nonlinear regime requires a model that associates
such criteria to objects.

\subsection{Spherical evolution and SVdW  model}
\label{sect:SVdW}

The spherical evolution model  provides a complete
description of the nonlinear evolution of a spherically
symmetric top-hat density perturbation. One of the main 
features of this model is that the evolution
does not depend on the initial size of the region, i.e.\ on
the initial radius or enclosed mass, but only on the amplitude of the initial
top-hat overdensity. 

For the collapse of perturbations, the spherical evolution model
in combination with the excursion set provides a good description of
the statistics of dark matter halos.  As we review in Appendix~\ref{sect:sphere},
collapse   occurs when the linear density fluctuation
reaches a critical value or barrier $\deltac$.   We can then use the excursion set formalism
to determine the fraction of trajectories that cross this barrier for the first time, accounting for the cloud-in-cloud process, within some $d\ln\sigma$
of a smoothing scale $\sigma$ through the differential fraction
\ba
f_{\ln\sigma}(\sigma) \equiv  \frac{{\rm d} f}{{\rm d }\ln\sigma}= \sqrt{\frac{2}{\pi}} \frac{\deltac}{\sigma} e^{-\frac{\delta^2_{\rm c}}{2\sigma^2}} \,.
\ea
Since both mass and number are conserved in the collapse, the linear theory mapping $\sigma(M)$ carries
over to the nonlinear regime and so the mass function, or the comoving differential number
density of halos is
\ba
\frac{{\rm d}n}{{\rm d}\ln M} = \frac{\rhom}{M} f_{\ln \sigma}(\sigma) \frac{{\rm d }\ln\sigma^{-1}}{{\rm d }\ln M},
\ea
where $\rhom/M$ is the number density of such objects if the fraction were unity.

We can extend the model to underdense regions in the initial density field.   These are naturally associated with
voids in the evolved density field today. A key assumption in making the
connection between the excursion set and the abundance of nonlinear objects is
that each collapse occurs in isolation. This makes sense for collapsing objects
since the comoving volume occupied shrinks. In contrast to overdense regions
which contract, voids expand. We shall see that this causes a problem for
mapping excursion set predictions onto the statistics of voids.

Nonetheless let us start with the simple spherical evolution model following
\cite{2004MNRAS.350..517S}.   The critical density threshold is
 defined to be when the expanding shells cross  \citep[see e.g.][]{1984PThPh..71..938S, 1984ApJ...281....9F, 1985ApJS...58....1B}.    
As shown in Appendix \ref{sect:sphere}
for an Einstein de-Sitter (EdS) universe, this
 occurs when the nonlinear average density  within the void reaches $\rhov = 0.2\rho_{\rm m}$ or when the linear density threshold reaches
  $\delta_{\rm v} = -2.7$. Note we will use this notation of $\rhov$ to refer to the nonlinear density of the void region and 
$\delta_{\rm v}$ to refer to the linear underdensity used as a threshold in the excursion set model.  We show in Appendix \ref{sect:sphere}, that these EdS values suffice for the
accuracy to which we wish to describe alternate cosmologies such as the $\Lambda$CDM model.

Once we have this value for the void barrier we can follow the excursion set formalism for
determining the fraction of random walks which pierce the barrier
$\delta_{\rm v}$.  Similar to the cloud-in-cloud process, the void-in-void process
accounts for the fact that a void of a given size may be embedded in another 
underdense region on a larger scale.  We thus define the first crossing distribution by 
associating the random walks with the smoothing scale for which they first
cross
the barrier $\delta_{\rm v}$.  

The second process, the void-in-cloud scenario, occurs when a void of a 
given size  is embedded in an overdense region on a larger scale, which will eventually collapse to a halo and squash the void out of existence.
In order to account for the void-in-cloud effect, \citet{2004MNRAS.350..517S} proposed that the excursion set method applied to voids requires a second barrier, the threshold for collapse of overdense regions,
$\deltac$. In calculating the first crossing distribution, \citet{2004MNRAS.350..517S} argued that we  need to 
determine the largest scale at which a trajectory crosses the barrier $\delta_{\rm v}$ given that it has not 
crossed $\deltac$ on any larger scale. They posit that  the value of $\deltac$ should lie somewhere in between
$\deltac =1.06$, the value at turnaround in the spherical collapse model, and $\deltac =1.686$, the value at 
the point of collapse \citep[see also][]{2012MNRAS.420.1648P}.
In this paper we shall refer to the  model of \citet{2004MNRAS.350..517S} as the \lq SVdW\rq\  model.

\begin{figure}
\centering
{\epsfxsize=8.truecm
\epsfbox{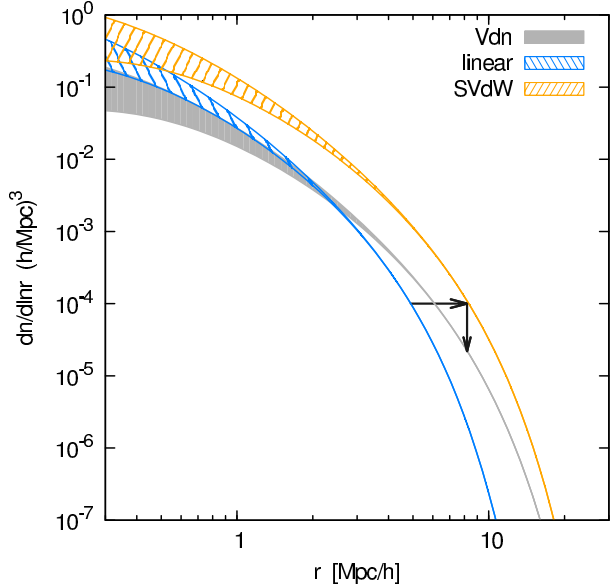}}
\caption{Void abundance model predictions. In the SVdW model, the number density
of linear underdensities  (blue  curve) remains unchanged in void formation
and only their sizes change (arrow to orange  curve). 
In
the \Vdn\ model, the number density also changes so as to conserve the volume
fraction in voids, lowering the amplitude at fixed shape (arrow to 
grey curve). Varying 
$1.06 \le \deltac \le 1.686$ (shaded or hatched regions) changes the abundance significantly only for small voids $r \la 1 h^{-1}$Mpc.   We take
$\delta_\text{v}=-2.7$ throughout. We use the $\sigma_8=0.8$ $\Lambda$CDM cosmology as listed in
Table \ref{table:simulations} here and in the following figures unless otherwise stated. 
}
\label{fig:2mappings}
\end{figure}

By the same reasoning as applied to halos,
the  SVdW formula for the abundance of voids in linear theory is given by 
\ba
\frac{{\rm d} n_{\rm L} }{{\rm d }\ln M} = \frac{\rhom}{M}  f_{\ln \sigma}(\sigma) \frac{{\rm d }\ln\sigma^{-1}}{{\rm d }\ln M} \, ,
\ea
where 
\ba
\label{eqn:exactf}
    f_{\ln \sigma}(\sigma)
    &=&2 \sum_{j=1}^\infty e^{-\frac{(j \pi x)^2}{2}} j\pi x^2 \sin(j\pi \mathcal{D}) ,
\ea
with
\ba
\mathcal{D} = \frac{|\delta_{\rm v}|}{\deltac + |\delta_{\rm v}|} \, ,\quad x =\frac{ \mathcal{D}  }{|\deltav|}\sigma\,.
\ea
Note that \citet{2004MNRAS.350..517S} give $f_{\ln S} = S df/dS = f_{\ln \sigma}/2$.
We have added the subscript ``L" to remind the reader that the logic relies on equating
a number density derived from linear theory to the number density of some nonlinear object 
for reasons that will be clear below.

Since the infinite series in equation (\ref{eqn:exactf}) is cumbersome to work with, it
is useful to have an accurate closed form expression.  As we discuss in Appendix
\ref{sect:mod}, the accuracy of the approximation given in  \citet{2004MNRAS.350..517S} is uncontrolled as $\sigma \rightarrow \infty$.   Instead we find  the limiting
forms for equation \eqref{eqn:exactf} such that the domain of validity of the approximation
is well-defined.    Note that as $\sigma \rightarrow 0$, the opposing barriers are high and  the sum must return the single barrier
expression since the probability of first crossing the collapse barrier is vanishingly small.
This fixes the form as $x\rightarrow 0$.   As $\mathcal{D}$ increases toward unity, we lower the collapse barrier relative to the void barrier and the value of $x$ at which this limit is approached decreases.  Correspondingly to achieve a matching at this point, we need to keep more terms in the sum.   The largest value that we will be interested in is
$\mathcal{D}<3/4$ and so it suffices to keep 4 terms
\ba
    f_{\ln \sigma}(\sigma)
\label{eqn:approxf}
    &\approx&
       \begin{cases}
     \sqrt{\dfrac{2}{\pi}}\dfrac{|\deltav|}{\sigma} e^{-\frac{\deltav^2}{2\sigma^2} },&x\le 0.276 \\
     2 \sum_{j=1}^4 e^{-\frac{(j \pi x)^2}{2}} j\pi x^2 \sin(j\pi \mathcal{D}), & x>0.276\\
     \end{cases}
\ea
which is accurate at the $0.2\%$ level or better across the domain of validity.
The approximation of equation\ \eqref{eqn:approxf} is used in all the numerical work throughout the paper.  

We can alternately express the number density in terms of the linear theory radius
of the void $\rL$.   Using $\rhom/M = 1/V(\rL)$ and defining the volume of a spherical
region of an arbitrary radius, $R$, as
\ba
V(R) \equiv \frac{4}{3} \pi R^3 \,,  
\ea
we obtain
\ba
\frac{{\rm d} n_{\rm L} }{{\rm d }\ln \rL} = \frac{ f_{\ln \sigma}(\sigma) }{V(\rL)} \frac{{\rm d }\ln\sigma^{-1}}{{\rm d }\ln\rL} \, .
\label{eqn:sw}
\ea

In the spherical evolution model, the actual void expands from its linear radius.   
At the epoch of shell crossing $\rhov = 0.2\rho_{\rm m}$.  Given that
\ba
\frac{\rNL}{\rL} = \left( \frac{\rhom}{\rhov} \right)^{1/3} \,,
\ea
spherical expansion predicts that this expansion factor is $\rNL 
\approx 1.7 \rL $.    The void abundance therefore becomes
\ba
\frac{{\rm d} n }{{\rm d }\ln \rNL} = \frac{{\rm d} n_{\rm L} }{{\rm d }\ln \rL} \Big|_{\rL = \rNL/1.7}\,.
 \label{eqn:swnl}
\ea
Note that in this model ${\rm d} n/{\rm d}\ln r$ shifts left to right in scale
through the nonlinear growth but does not change in amplitude, as is shown in
Fig. \ref{fig:2mappings}. 

The SVdW model has two parameters $\deltac$ and $\deltav$.     The latter is fixed
by the shell-crossing criterion whereas the former is expected to vary within
$1.06 \le \deltac \le 1.686$.
In Fig. \ref{fig:2mappings}, we also show that for the range of radius of interest
($r>1h^{-1}\text{Mpc}$), changing $\deltac$ within its expected range has little effect on the
void abundance.   

\begin{figure}
\centering
{\epsfxsize=8.truecm
\epsfbox{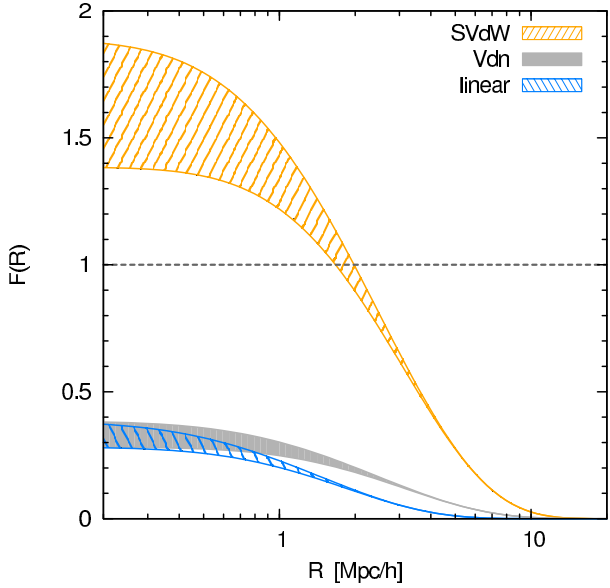}}
\caption{The cumulative volume fraction in voids with radii larger than $R$ for
the various models: linear theory (blue striped region, $R=\rL$), SVdW model
(orange striped region, $R=\rNL$), \Vdn\ model (grey shaded region, $R=\rNL$).
Regions correspond to the expected range of $1.06 \le \deltac \le 1.686$ and we take $\delta_\text{v}=-2.7$
throughout. For SVdW the fraction unphysically exceeds unity at $R \approx 2
h^{-1}\text{Mpc}$ while for the \Vdn\ model conserves the total fraction from
the linear theory of $\mathcal{F}(0)\approx 0.3$. 
}
\label{fig:volfrac}
\end{figure}

The SVdW model makes a very specific prediction for the abundance of large
voids. Again the key assumption of the SVdW model is that the comoving number density
of objects is conserved during the evolution $n=n_{\rm L}$ and only their size
has changed. Unfortunately, for spherical evolution this assumption is invalid
for large voids. In particular, the cumulative volume fraction in voids larger
than $R$ defined as
\ba
    \mathcal{F}(R) = \int_R^\infty \frac{{\rm d }\rNL}{\rNL} \, V(\rNL) \frac{{\rm d}n}{{\rm d}\ln \rNL},
\ea
exceeds unity for radii of interest. In Fig. \ref{fig:volfrac}, we demonstrate
that this problem cannot be cured by changing $\deltac$ within the expected
range as it only affects small voids whereas the problem appears at $R \approx 2
h^{-1}\text{Mpc}$.  Indeed if we take $R \rightarrow 0$ then for
the exact $f_{\ln\sigma}$ given by equation \eqref{eqn:exactf} \citep{2004MNRAS.350..517S}
\ba
\label{eqn:totvol}
\mathcal{F}(0) = \left( \frac{\rNL}{\rL} \right)^3 \int_0^\infty \frac{d \sigma}{\sigma} f_{\ln \sigma} = \left( \frac{\rNL}{\rL} \right)^3 (1-\mathcal{D}).
\ea

\begin{figure}
\centering
{\epsfxsize=8.truecm
\epsfbox{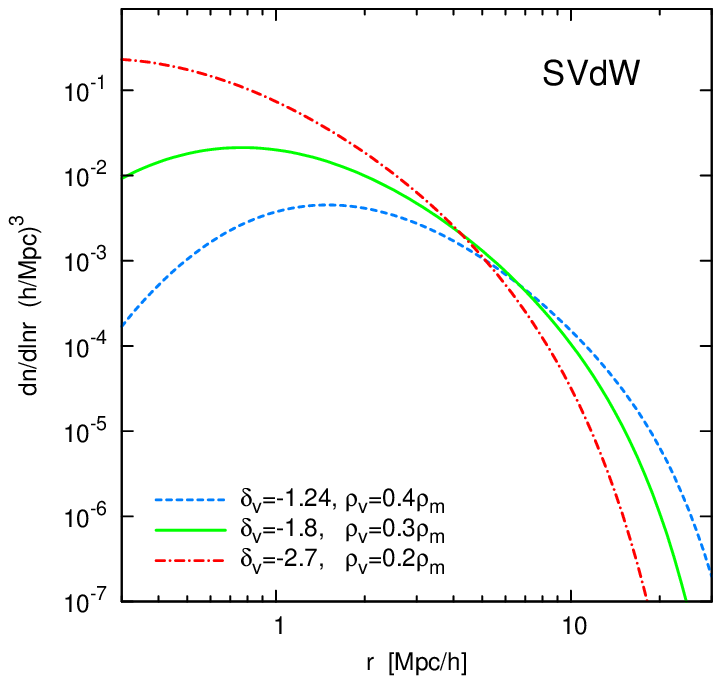}}
\vskip 0.25cm
{\epsfxsize=8.truecm
\epsfbox{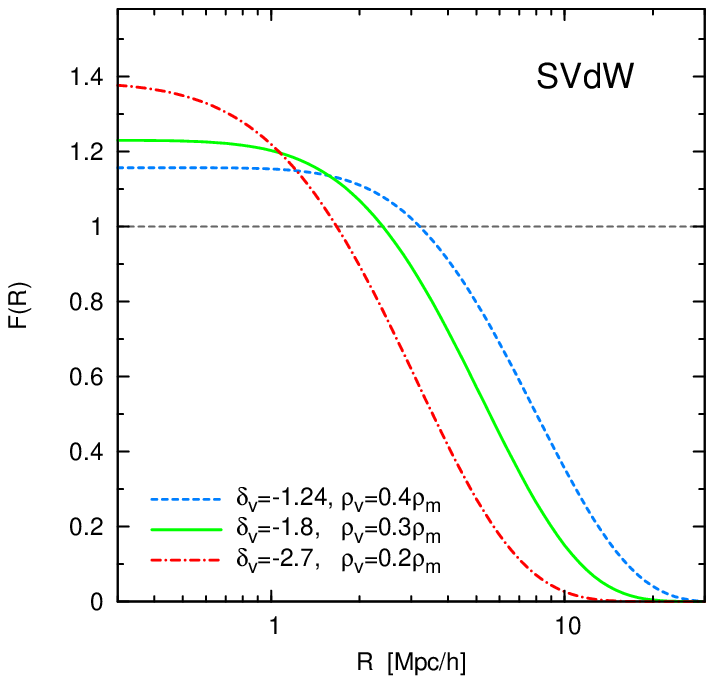}}
\caption{Relaxing the shell crossing criteria of the SVdW model void abundance predictions.
\emph{Upper:} Variation of the void 
underdensity $\rhov/\rhom$  changes both the shape of the abundance through the linear barrier
$\delta_\text{v}$ and the size of the voids or horizontal shift through
 $\rNL/\rL = (\rhov/\rhom)^{-1/3}$.  Decreasing $|\delta_\text{v}|$ increases the number of large voids and decreases that of small voids. 
\emph{Lower:} The cumulative volume fraction  in voids with radii larger than $R$
decreases as $|\deltav| \rightarrow 0$ and $R \rightarrow 0$
but at the expense of making the larger voids more abundant. 
We use $\delta_\text{c}=1.06$ throughout. }
\label{fig:deltav}
\end{figure}

This result suggests that reducing $\deltav \rightarrow 0$ simultaneously takes
 $\rNL \rightarrow \rL$ and 
$\mathcal{D} \rightarrow 0$ bringing SVdW asymptotically to physicality $\mathcal{F}(0)=1$.
 Strictly speaking,  $\deltav$ is  fixed by the shell-crossing
criterion.  However, given the approximate nature of the correspondence between the isolated spherical
expansion model and real voids, it is interesting to explore whether  modifications
to this criterion can bring the SVdW model into agreement with physicality and simulations.    If we change the nonlinear density at which voids are defined
$\rhov/\rhom$, the linear density threshold $\deltav$ and the expansion factor
$\rNL/\rL$ must change in a self-consistent fashion (see Fig.~\ref{fig:alwaysfillup}
and equation \eqref{eqn:deltavrNL}).   
In Fig. \ref{fig:deltav} we show
that changing $\delta_\text{v}$ alters the shape of the abundance function.   As $|\deltav|$ decreases, the steepness of the
abundance function also decreases.   
Thus, although lowering $\deltav$ can make the total volume fraction physical (Fig.~\ref{fig:deltav}, lower panel), it
increases the abundance of the largest voids.   
We shall see that
the agreement between simulations and the abundance of voids in the 
excursion set method is remarkably good if we remove the assumption of isolated spherical expansion.
Assuming that the number density of voids is conserved as they expand causes the SVdW model to greatly
overpredict the abundance
of large voids regardless of the choice of $\deltac$ and $\deltav$.

\subsection{Volume conserving \Vdn\ model}
\label{sect:Vdn}

We propose a simple fix to the unphysicality of the isolated spherical expansion model for voids.
  We require that the volume fraction and shape of the abundance function is 
fixed during the expansion, rather than assuming that the expansion of isolated voids preserves their total number density.  
Specifically, if
we define the volume fraction in linear theory, $\mathcal{F}_{\rm L}$, as
\ba \label{eqn:FL}
\mathcal{F}_{\rm L}(R_{\rm L}) = \int_{R_{\rm L}}^\infty \frac{  {\rm d } \rL }{\rL} V(\rL)  \frac{{\rm d}n_{\rm L}}{{\rm d} \ln \rL}\,.
\ea
then this fraction is conserved if we define the nonlinear abundance as
\ba
\label{eqn:volfrac_conserv}
V(\rNL) {\rm d}n  = V(\rL) {\rm d}n_{\rm L}\big|_{\rL(\rNL)}.
\ea
In this picture, when 
a void expands from $\rL \rightarrow \rNL$ it combines with its neighbours to conserve
volume and not number.   
Thus the abundance becomes
\ba
\frac{{\rm d} n }{{\rm d }\ln \rNL} &=&  \frac{ V(\rL) }{V(\rNL)} 
\frac{{\rm d} n_{\rm L}}{{\rm d }\ln \rL}
\frac{{\rm d} \ln \rL}{{\rm d }\ln \rNL} 
 \Big|_{\rL (\rNL)} \nonumber\\
 &=&  \frac{ f_{\ln \sigma}(\sigma)}{V(r)}  \frac{{\rm d }\ln\sigma^{-1}}{{\rm d }\ln\rL}
 \frac{{\rm d} \ln \rL}{{\rm d }\ln \rNL} 
 \Big|_{\rL (\rNL)}
\,.
 \label{eqn:Vdn}
\ea
We call this model the \Vdn\ model and show its abundance prediction in Fig.
\ref{fig:2mappings}. We have left the mapping $\rNL(\rL)$ general here since the
specific form from isolated spherical expansion until shell crossing may not
apply here. We will however adopt $\rNL=1.7\rL$ for voids with nonlinear density
$\rhov = 0.2\rho_{\rm m}$ from N-body simulations as a starting point.
Note that in this case $ {{\rm d} \ln \rL}/{{\rm d }\ln \rNL} =1$ and the impact
of going from the linear to the nonlinear abundance is both a shift in scale and
a change in amplitude with no change in shape, as is shown as the combination of arrows in
Fig.\ \ref{fig:2mappings}.

In Fig.\ \ref{fig:volfrac}, we also show the cumulative volume fraction with this abundance
function, along with that for linear theory defined in
equation \eqref{eqn:FL}.  Since by construction the volume fraction is conserved,
the two curves differ only by a horizontal shift in scale.

\begin{figure}
\centering
{\epsfxsize=8.truecm
\epsfbox{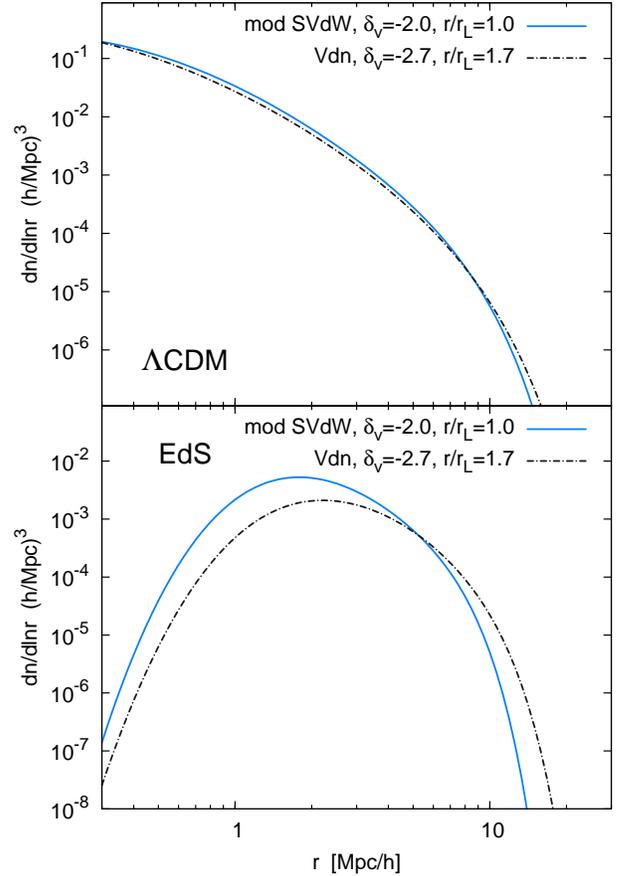}}
\caption{Void abundance in the \Vdn\ model (black dot-dashed curve)
and a modified SVdW model (blue solid line)
with ad hoc variations 
designed to fit the $\Lambda$CDM  \Vdn\ model. Upper panel: we choose $\delta_\text{v}=-2$ and
$r/r_\text{L}=1$ in the SVdW model in violation of spherical expansion predictions in a
$\sigma_8=0.8$ $\Lambda$CDM cosmology.
Lower panel: we show that the same set of parameters give a poor fit,
in an EdS model (see Table
\ref{table:simulations}). 
For all curves we use $\delta_\text{c}=1.686$.
}
\label{fig:nouniversal}
\end{figure}

Since the \Vdn\ model is not the unique means of constructing a physical model, it is interesting to explore other ways of keeping the volume fraction below unity. 
Phenomenologically, we can decouple the relationship in equation~\eqref{eqn:deltavrNL}
between the
parameters  $\delta_\text{v}$ and $r/r_\text{L}$ provided by the spherical
expansion model.   In fact, we can choose these parameters so as to mimic the
\Vdn\ predictions for a fixed cosmology.  For example, in the upper panel of Fig. \ref{fig:nouniversal}, we
find we can change the parameters $\delta_\text{v}\to-2$ and $r/r_\text{L}\to1$
in the SVdW model to fit the \Vdn\ model in the $\sigma_8=0.8$ $\Lambda$CDM
cosmology listed in Table \ref{table:simulations}. However, this change
then predicts very different abundances than the \Vdn\ model for a different
cosmology as shown with the EdS
cosmology listed in Table \ref{table:simulations} and Fig.~\ref{fig:nouniversal} (lower panel).
We shall show below that simulation results favor the \Vdn\ model over universal
changes in $\delta_\text{v}$ and $r/r_\text{L}$.
The \Vdn\ model retains the $\delta_{\rm c}$ parameter from the SVdW model to 
describe the influence of surrounding mass concentrations on the growth of voids. 
Within the excursion set method this
parameter accounts for the crushing of small voids which reside in over dense regions. 
Note that the assumption of spherical expansion of these small voids, as 
well as the spherical collapse of the larger overdense region should be 
taken as a simple approximation which will not be accurate for 
small non-spherical voids. In this study we focus on testing the model using simulations 
of voids with radii $r>1 h^{-1}$Mpc whose abundance is not affected by this crushing effect.

\section{Simple void finding algorithm }
\label{sect:void_finder}

In Section \ref{subsect:void_finder} we outline the  void finding algorithm used to identify 
voids in both the dark matter and halo populations in this work.
In Section \ref{subsect:nbody} we present the details of the N-body simulations which were carried out as summarised in Table \ref{table:simulations}.

\subsection{Void finder}

\label{subsect:void_finder}
As we have already mentioned 
one of the main complications in studying the distribution of voids in the large scale structure of the Universe,
is  finding a robust definition of what a void is \citep[see][for a comparison of void finders]{2008MNRAS.387..933C}. In this work we wish to make a direct comparison
 to the predictions of the excursion set formalism which
assumes that these underdense regions are non-overlapping spheres of a given underdensity corresponding to a region at the moment 
of shell crossing. We shall retain  the moment of shell crossing as the key feature which defines a 
nonlinear void in the matter distribution today although we also compare the measured abundance of voids with
different underdensities to the predictions of the volume conserving model in Section \ref{subsect:comparing_vdn}. 

We start with the publicly available
 code  {\sc ZOBOV} \citep{2008MNRAS.386.2101N} which uses Voronoi tessellation to estimate densities and
find both voids and subvoids.  
The main advantage of using tessellation methods is that it gives a local density estimate by dividing space into cells,
where the cell around any given particle is the
region of space closer to that particle than to any other. The Voronoi tessellation also
gives a natural set of neighbours for each particle which {\sc ZOBOV} uses to construct zones around density minima.
 
The output from ZOBOV is useful  for our purposes for two main reasons.
Firstly, it outputs a linked list of zones in the dark matter distribution, which is also ordered by density contrast.  It
  thus provides a tree structure which 
we can prune according to the definition of a void.
Secondly, {\sc ZOBOV} identifies the \lq core\rq\  or least dense particle in a zone and returns its density as well
as a measure of the probabilities that each collection of zones arises from Poisson fluctuations.
Note that the list which {\sc ZOBOV} returns contains zones of various densities and Poisson probabilities,  
some of which could be overdense or not statistically significant, so it is necessary to prune the output from {\sc ZOBOV} in order to construct a  void catalogue.

In constructing our void finder the goal is to identify all spherical non-overlapping underdense regions
of average density $\rhov=0.2\rhom$ in a dark matter simulation. We use
the output from {\sc ZOBOV} and find spherical regions  centred around the core particle (lowest density particle) in
a zone, which can encompasses any particles which are around the zone returned by {\sc ZOBOV}  and not necessarily 
part of the particular zone or collection of
zones returned by {\sc ZOBOV}. 
\begin{table}
\caption{A toy example showing the first five columns  from a {\sc ZOBOV} output file.
}
\centering
\begin{tabular}{cccccccc  }
\small{Void\#(zones)} & \small{FileVoid\#}& \small{CorePar} &\small{CoreDen} &\small{ZoneVol} \\
1 ($a,b,c$) &2945 &26 &1.08e-02 &4.9e+03  \\
2 ($b,c$) &5033 &83 &1.8e-02 &7.8e+02 \\
3 ($c$)&1814 &45 &2.0e-01 &1.9e+02  \\
\end{tabular}
\label{table:zobov}
\end{table}

\begin{figure}
\centering
{\epsfxsize=5.5truecm
\epsfbox[0 0 280 320]{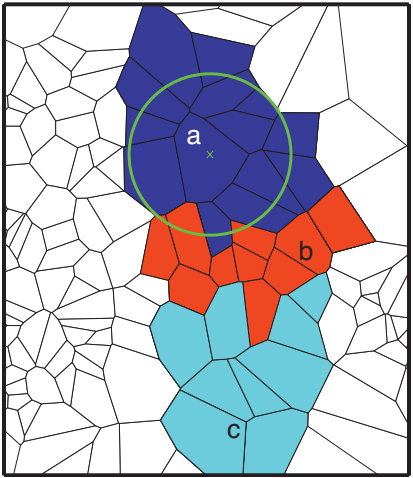}}
\caption{An illustration of a spherical void identified using the zones output from {\sc ZOBOV} after Voronoi tessellation of the region.
The Void \# 1 output from {\sc ZOBOV} given in Table \ref{table:zobov}  is shown as three shaded zones, {\it a} (blue), { \it b} (red) and {\it c} (cyan).
The core particle of zone {\it a} is shown as a green cross while the void we identify in this region of a given density is shown as a green circle.
}
\label{fig:void_finder}
\end{figure}

\begin{table*}
\begin{minipage}{123mm}
\caption{Details of the simulations used in this work.}
\begin{tabular}{@{}lccccccc}
\hline\hline
Model & $\Omega_{\rm m}$ & $h$ & $\sigma_8$ & Boxsize & \# particles & \# simulations & $z$ output \\
\hline
$\Lambda$CDM & 0.26 & 0.715 & 0.8 & 64, 128, 256 & 256$^3$ & 8 & 0, 0.5, 1 \\
 &  &  &  & 500 & 400$^3$ & 1 & 0  \\
 &  &  &  & 500 & 256$^3$ & 8 & 0 \\
\hline
$\Lambda$CDM & 0.26 & 0.715 & 0.9 & 64, 128, 256 & 256$^3$ & 8 & 0, 0.5, 1 \\
\hline
EdS & 1 & 0.7 & 0.8 & 64, 128, 256 & 256$^3$ & 8 & 0 \\
\end{tabular}
\label{table:simulations}
\end{minipage}
\end{table*}

One of the outputs from {\sc ZOBOV} is a text file which lists individual zones and
joined zones which are added to the list in a sequential process
analogous to water flooding a plane with troughs of various heights \citep[see also][]{2007MNRAS.380..551P}. During flooding, when water
from a particular zone  or joined zones flows into a neighbouring deeper zone, the process stops and the zone
is recorded in the list. A toy example of the output is shown in Table \ref{table:zobov}
where the zones are listed in order of density contrast.  
Fig. \ref{fig:void_finder} shows an illustration of the Voronoi tessellation of the region surrounding Void \# 1 output from {\sc ZOBOV} and given in Table \ref{table:zobov} which is made up of zones {\it a} (blue), {\it b} (red) and {\it c} (cyan). 
In this figure the core particle (CorePar = 26) of zone {\it a} is shown as a green cross.
Here FileVoid\# and CorePar refer to unique identification tags for the void and its core particle respectively.
CoreDen is the density, in units of the mean, of the void's core particle.

 In order to count non-overlapping regions with an average density of 0.2, we perform
the following two stages of analysis on the output from {\sc ZOBOV}. Firstly, starting from the
top of the ZOBOV output file we determine if the first collection of zones listed e.g Void \#1,
 which is made up of zones {\it a}, {\it b} and {\it c},
 pass the following criteria:
\begin{itemize}
\item $[r_{\text{min}},r_{\text{max}}]$: The radius corresponding to a sphere of equal volume should be  $>r_{\text{min}}$ and $<r_{\text{max}}$.
\item The core particle density is $<0.2 \rhom$.
\end{itemize}
As we are searching for regions which have an average 
density $\rhov = 0.2 \rho_{\rm m}$ we also only consider zones in the list which have a density 
$\rhov \ge 0.2 \rho_{\rm m}$ in order to speed up the search. 

If the collection of zones fulfils all of the above, then we proceed to the second stage. 
A spherical region around the centre is found
by iteratively including one
particle at a time moving away from the centre of the void until $\rhov=0.2\rhom$.
We assume that the volume corresponds to a sphere with radius equal to
the distance from the centre to the last particle included. 
If for example, Void \#1 did not pass the $r_{\text{max}},r_{\text{min}}$ criteria
 then we consider if the deepest zone, zone {\it a}, does and 
if so we grow a sphere around the centre of this single zone.
This step is important as the output from {\sc ZOBOV} only lists the deepest zone \lq {\it a}\rq \, once and if Void \#1 fails the 
$r_{\text{max}},r_{\text{min}}$ criteria the void finder would miss counting 
the deepest zone in the simulation box.
In Fig. \ref{fig:void_finder} we show an illustration of the spherical void (green circle) which is grown around the core particle (green cross) for Void \# 1. The spherical region is not necessarily  restricted to include particles in the zones returned by {\it ZOBOV}.

We then proceed to the next line in the output file 
and perform the same two stages of analysis.  
All of the particles in the spherical regions which are grown have been tagged and at any stage if there is any 
overlap of spheres we disregard the less underdense zone to avoid double counting any volume in the simulation.
In Fig. \ref{fig:void_finder} this corresponds to growing another sphere around the core particle in zone {\it b}, until the required density is reached. This spherical void is added to the  catalogue if it does not overlap with the void around zone {\it a} (green circle).
Note if we consider the output from {\sc ZOBOV} as a tree structure then this procedure is similar to 
walking the tree from root to tip, pruning any branches after our criteria are met.

Using a cut in $r_{\rm max}$ and $r_{\rm min}$ as above allows us to avoid considering spuriously small voids and the first output in the text file which 
is a void which takes up the entire simulation box, however we have checked that our results are not sensitive to changes in 
$r_{\text{max}}$ and $r_{\text{min}}$ but we retain these criteria in the void finder to speed up computation.
For our simulations we use the following: $[r_{\text{min}},r_{\text{max}}] = [0.5,15], [1,30]$, $ [2,60]$ and [4,120] $h^{-1}$Mpc 
for the $64$$h^{-1}$Mpc, $128$$h^{-1}$Mpc and $256$$h^{-1}$Mpc and $500$$h^{-1}$Mpc boxes respectively.

We tested several different criteria in identifying voids and found that the two points listed above are sufficient to identify significant non-overlapping regions of a given 
underdensity in the particle distribution. In testing the robustness of our void finder we considered the impact of the following adjustments to the method:
\begin{itemize}
\item As an alternative to using the core particle to define the centre of the void we can use a volume-weighted centre defined as
\begin{eqnarray}
\vec{c} = \frac{\sum_i \vec{x}_iV_i}{\sum_i V_i},
\end{eqnarray}
where $\vec{x_i}$ and $V_i$ are the position and volume of each particle in the zones returned by ZOBOV respectively. 
We found that the abundance of voids was not substantially affected by this choice and so we use 
the core particle as the centre of the spherical region.
Note that using the volume-weighted centre is more robust when using stacked voids \citep[see e.g.][]{2012ApJ...754..109L}.
\item Instead of allowing the spherical region to include all particles which are around the void we consider restricting it so that only
particles which {\sc ZOBOV} list as being part of the zone are  included when growing the sphere.
 We find that in the majority of cases the region of underdensity $\rhov = 0.2 \rho_{\rm m}$ is contained within the collection of zones returned by {\sc ZOBOV}
and so this restriction does not affect the measured abundance of voids.
Our results in Section \ref{sect:results} use all particles within 1.5 times the radius of the zones to find the spherical void.
Including particles only within this radius was found to be sufficient considering the original collection of zones was required
to have an average density of $>0.2\rho_{\rm m}$.
An alternative approach to this would be to use the actual volume of each zone particle when trying to find a
void of a given average density. This would allow for irregularly shaped voids which it could be argued is a more \lq natural\rq \, description
of an actual void, however as we mentioned we are trying to compare with the excursion set theory for abundances which assumes spherical voids.
\item We originally included a third criteria in our void finder by requiring that the probability a zone arose from a 
Poisson process was less than a given significance \citep[see][for more details]{2008MNRAS.386.2101N}.
However in the context of our void finder, we found that the core particle density requirement by itself was sufficient  to get rid of spurious voids. This also  agrees with the findings of  \cite{2008MNRAS.386.2101N}.
\item In the algorithm we have described we stop growing a sphere around the core 
particle when we find the desired underdensity  at the maximum radius at which this 
occurs within the radius of the collection of zones. This is a different approach to simply stopping to record the first
radius where  $\rhov = 0.2 \rho_{\rm m}$ which would not take into account  void-in-void scenarios. 
In practice we find that accounting for a void-in-void effect alters the measured abundances by a small amount (e.g $\sim 7\%$ over the range $1<r(h^{-1}$Mpc$)<10$).
\item The above method does not allow any overlap of voids within the simulation in order to compare with the excursion set method.
In practise for regions of $\rhov = 0.2 \rho_{\rm m}$ we found the overlap was very small for the simulations we consider.
\item {\sc ZOBOV} is run using all the particles in the simulation with a run-time
density threshold parameter which can limit the growth of a collection of zones into high-density regions. We set this parameter to 0.2 however we have verified 
that changing this run time parameter has little effect on the abundance of underdensities found by our void finder.
\end{itemize}

\subsection{$N$-body simulations }
\label{subsect:nbody}

We measure the abundance of voids in the dark matter distribution using  a 
series of N-body simulations in various box sizes.
These simulations were carried out at the University of Chicago using the  TreePM
 simulation  code  {\tt Gadget-2} \citep{Springel:2005mi}. The $\Lambda$CDM model used has
the following cosmological parameters:
$\Omega_{\rm m} = 0.26$,
 $\Omega_{\rmn{DE}}=0.74$, $\Omega_{\rm b} = 0.044$,
$h = 0.715$ and a spectral tilt of $n_{\mbox{s}} =0.96$ \citep{Sanchez:2009jq}.
The  linear theory rms fluctuation
in spheres of radius 8 $h^{-1}$ Mpc is set to be  $\sigma_8 = 0.8$ for our main simulation set of 8 independent 
realisations of the $\Lambda$CDM cosmology.
In order to investigate the abundance of voids in different cosmologies we also carry out two additional simulations;
one with a $\Lambda$CDM cosmology and $\sigma_8 = 0.9$ and another which we refer to as the \lq EdS\rq \,simulation which 
has $\Omega_{\rm m} = 1$. The EdS simulation is not a viable cosmological model for our Universe as it has already been ruled out by many 
observations but we use it here as a tool to examine how robust our void models are
to large changes in the power spectrum or cosmology.

The simulation details are summarised in Table \ref{table:simulations}. Most of the simulations use $N=256^3$ 
particles to represent the dark matter while for the larger simulation box of $500 h^{-1}$Mpc we use 400$^3$ particles.
The error on the abundance of voids measured in the $500 h^{-1}$Mpc box is estimated from eight lower resolution simulations which have 
$256^3$ particles in a 
computational box of $500 h^{-1}$Mpc on a side. These lower resolution simulations have a mean abundance which agrees with the $400^3$ 
particle simulation over the range of scales which we consider and 
are computational less expensive to run and analyse with 
the void finder. 
The initial conditions of the particle load were set up with a
 glass configuration of particles \citep{Baugh:1995hv} and 
the Zeldovich approximation to displace the particles from their initial positions.
We chose a starting redshift of $z=100$ in order to limit the discreteness effects of the initial displacement scheme \citep{Smith:2002dz}.
The linear theory power spectrum used to generate the initial
conditions was created using the CAMB package of \citet{Lewis:2002ah}.
Snapshot outputs of the dark matter distribution as well as the group catalogues were made at redshifts  1, 0.5 and 0.  In the following section we also test voids that are identified with
dark matter halos.
The simulation code {\tt Gadget2} has an inbuilt friends-of-friends (FOF) halo finder which was applied
to produce  halo catalogues of dark matter particles with 10 or more particles. A linking length of 0.2 times the mean interparticle separation was used in the halo finder.

\begin{figure}
\centering
{\epsfxsize=8.truecm
\epsfbox[ 80 360 436 685]{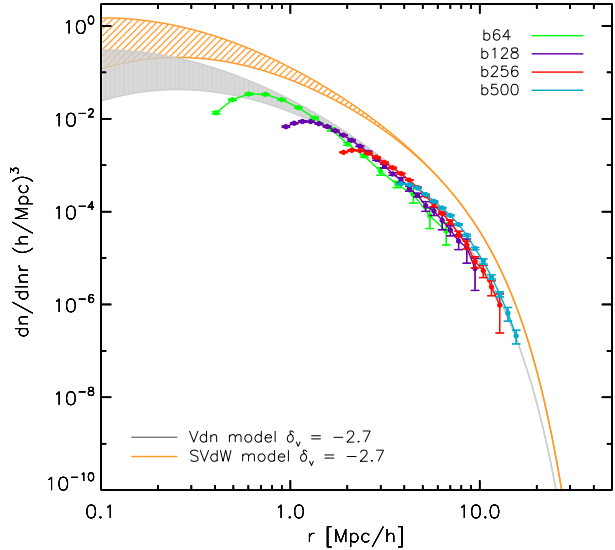}}
\caption{Void abundance in simulations vs.\ predictions  with $\rhov=0.2\rho_{\rm m}$
in the dark matter distribution of the $\sigma_8 = 0.8$ $\Lambda$CDM cosmology
in simulation box sizes 64$h^{-1}$Mpc (green),
128$h^{-1}$Mpc (purple), 256$h^{-1}$Mpc (red)  and 500$h^{-1}$Mpc (cyan) on a side. The error bars represent the scatter on the mean from eight different
realisations of this cosmology in each box size. The range in predictions 
cover the parameter interval $\deltac$ = [1.06,1.686] with
$\delta_{\rm v} = -2.7$ and are consistent with simulations for \Vdn\ 
(grey shaded) but not SVdW models (orange hatched).}
\label{64_128_256_500}
\end{figure}

\begin{figure*}
\centering
{\epsfxsize=16.5truecm
\epsfbox[ 56 373 547 578]{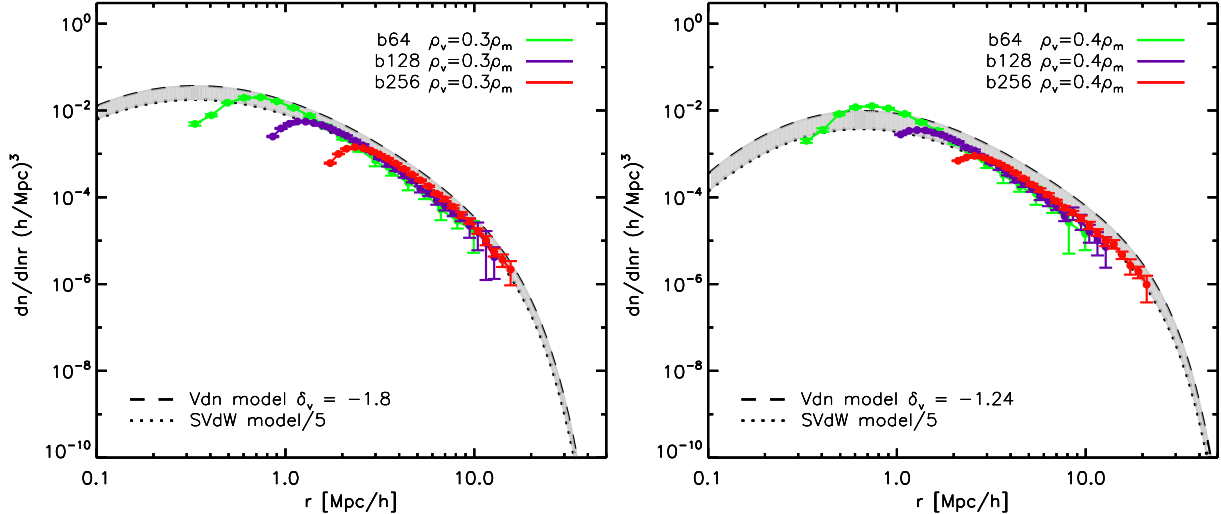}}
\caption{Void abundance for different defining underdensities  $\rhov = 0.3 \rhom$ (left panel), $\rhov = 0.4 \rhom$ (right panel).
The grey shaded region represents the excursion set predictions with varying amplitude and using a linear underdensity value
$\delta_{\rm c} = 1.686$ and
 $\delta_{\rm v}$,
given in the legend in each panel.  The amplitude rescaling vs the SVdW predictions
ranges from $\rhov/\rhom$ (\Vdn; top black dashed curve) to $1/5$ (bottom black dotted curve) which both preserve
agreement for $\rhov= 0.2\rhom$.
}
\label{rho_0.3}
\end{figure*}

\section{ Results}
 \label{sect:results}

In the following sections we compare simulation results for the abundance of voids with the predictions of both the
volume conserving and the SVdW model.
In Section \ref{subsect:comparing_vdn} we present the measured abundances of voids 
from the $\Lambda$CDM simulations in four different simulation box sizes.  
In Sections \ref{subsect:underdensity}, \ref{subsect:z_dep},  \ref{subsect:cosmology_dep}, we test the robustness of the models to variation in the critical
void underdensity, redshift, and cosmology respectively.  
In Section \ref{subsect:halo_voids} we present the abundance of voids identified 
in the dark matter halo catalogue.

\subsection{Baseline model  comparison}
\label{subsect:comparing_vdn}

We implement the void finder, which is described in Section \ref{subsect:void_finder}, to measure the abundance of spherical voids which have 
$\rhov = 0.2 \rho_{\rm m}$ at $z=0$ in all four simulation box sizes of the $\Lambda$CDM cosmology, see Table \ref{table:simulations}.
Fig. \ref{64_128_256_500} shows the average number density as a function of radius, of voids measured from
eight different realisations of the $\Lambda$CDM cosmology in simulation box sizes 64$h^{-1}$Mpc (green),
128$h^{-1}$Mpc (purple), 256$h^{-1}$Mpc (red) and 500$h^{-1}$Mpc (cyan) on a side. The error bars represent the scatter amongst these simulations.

The orange hatched region in this figure represents the SVdW model 
within the parameter interval $\deltac$ = [1.06,1.686] and
$\delta_{\rm v} = -2.7$
 and assuming that the 
voids have expanded by a factor of $1.7$ today. 
The grey shaded region  shows the \Vdn\ model for the same
parameters.   As discussed in Section \ref{sect:SVdW}, the range in $\deltac$
accounts for the void-in-cloud process by which a void in a larger
 overdense
regions  will be crushed  out of existence.  As we 
can see from Fig. \ref{64_128_256_500} this
only affects the smallest voids of $r<1$$h^{-1}$Mpc and for larger voids the abundance is insensitive to $\deltac$.
The decrease in the void abundance at $r (h^{-1} {\rm Mpc})\sim2.5, 1.5$ and $1$ for the $256$$h^{-1}$Mpc, 128$h^{-1}$Mpc and 64$h^{-1}$Mpc boxes shows the 
 resolution limit for each of these simulations where small voids are not fully  resolved and so the abundance is decreased.

As a result of assuming an isolated spherical expansion model the SVdW model overpredicts the abundance by a
factor of 5 whereas the \Vdn\ model agrees with simulations to  $\sim 16$\%  across the range $1<r(h^{-1}$Mpc$)<15$ where the results 
measured from simulations in different box sizes has converged.
This shows that the excusion set model is in good agreement with 
simulations once we account for the fact that voids merge as they expand and do not conserve the linear theory number density.
The \Vdn\ model conserves the volume rather than the number of voids and hence implies   that the number
density decreases in going from the linear to the nonlinear regime by the same amount that the volume of the voids grow.
It is somewhat surprising that using the factor of 1.7 in this model, 
which applies to the expansion of isolated objects, fits the results from the simulations where voids have merged as they expand.   It is important to test that this
is not just a coincidence but rather is robust to other choices of parameters in the
simulations.

\subsection{Underdensity variation}
\label{subsect:underdensity}

In both the SVdW and \Vdn\ models, we adopt the shell crossing criteria $\rhov= 0.2 \rhom$
for defining the void and match predictions to $\rhov$ as defined by the simple void finder of
Section \ref{subsect:void_finder}.   If the agreement between the \Vdn\ model and simulations
was robust, we would expect that it would be preserved for at least small variations in this
criteria.

We modify our  void
finder such that the largest non-overlapping spherical regions 
which have densities $\rhov = 0.3 \rho_{\rm m}$ and $\rhov = 0.4 \rho_{\rm m}$ are recovered from 
the simulations.  The results are shown in the left and right 
panels of Fig. \ref{rho_0.3} respectively. The errors plotted in this figure represent the scatter on the mean from eight simulations.

As discussed in Section \ref{sect:SVdW}  (see also Fig.~\ref{fig:alwaysfillup}
and equation~\eqref{eqn:deltavrNL}), changing the underdensity criteria in the spherical
evolution model alters  the shape of the abundance function through the linear threshold
$\deltav$.    Specifically, for $\rhov=0.3\rhom$,
$\deltav = -1.8$; while for $\rhov=0.4\rhom$,
$\deltav = -1.24$.

For dark matter voids with $\rhov = 0.2 \rhom$ the predicted abundance for the \Vdn\ model are approximately 
a factor of 5 smaller then those of the SVdW model. 
In modelling the number density of underdense regions with $\rhov = 0.3 \rhom$ and $\rhov = 0.4 \rho_{\rm m}$,
 which cannot be directly
related to shell crossing in the spherical expansion model, we adopt a phenomenological approach.   Given that for $\rhov = 0.2 \rhom$, the \Vdn\ model has the same
shape as the SVdW model but  a factor of $\rhov/\rhom=1/5$ lower amplitude, we can
preserve the good match there by either following the \Vdn\ prescription literally    and rescaling  SVdW by $\rhov/\rhom$ or 
by simply keeping this factor fixed at 1/5.

This range is plotted in Fig. \ref{rho_0.3} as grey shaded regions bounded by the two
limiting cases, black dashed and dotted lines.
Simulation results clearly favor the simple phenomenological prescription of rescaling
the amplitude by 1/5.   The $\rhov/\rhom$ scaling prescription of the \Vdn\ model
would overpredict the amplitude by 
 approximately 1.5 for voids with $\rhov= 0.3\rhom$  and 2 
 for 
$\rhov = 0.4\rhom$.   These results again highlight the point that excursion set models
predict the overall shape of abundance function accurately and only the amplitude needs to be altered 
to fit the simulations results, here without the benefit of volume conservation as motivation.  Note that the preferred rescaling of 1/5 is more than sufficient to bring the predictions to a physical
void filling fraction for $\rhov\ge 0.2\rhom$.

\subsection{Redshift variation \label{subsect:z_dep}}

\begin{figure}
\centering
{\epsfxsize=8.truecm
\epsfbox[ 80 360 436 685]{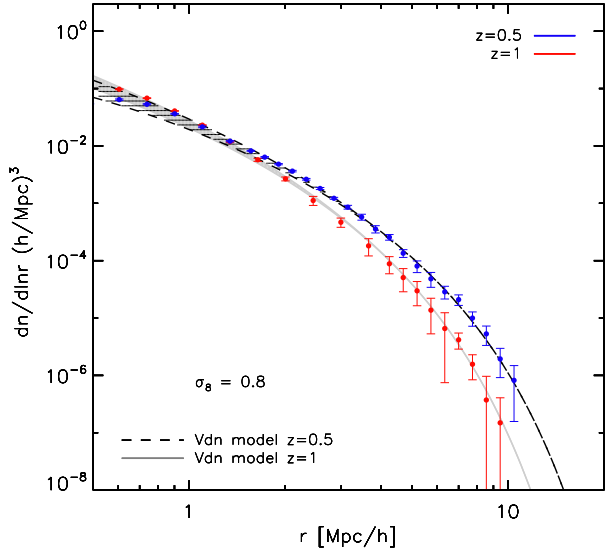}}
\caption{Redshift dependence of the void abundance with $\rhov = 0.2\rhom $ at $z=0.5$ (blue) and $z=1$ (red) measured from
the $\Lambda$CDM, $\sigma_8 = 0.8$  simulation.
The black hatched  (grey shaded) region represent the \Vdn\ model using linear underdensity values of
$\deltav = -2.7$ at $z=0.5$ ($z=1$) for the range $\deltac = [1.06,1.686]$.
Note the measured average abundances and errors
from the $256$$h^{-1}$Mpc, 128$h^{-1}$Mpc and 64$h^{-1}$Mpc  simulation boxes are the volume-weighted values.
Note in this figure we have plotted the results from the three simulation boxes using the same colour for clarity.
}
\label{fig:z_dep}
\end{figure}

Next we check the robustness of results to the redshift at which the void abundance
is measured.
Fig. \ref{fig:z_dep} shows the number density of voids as a function of radius at $z=0.5$ (blue) and $z=1$ (red)
measured from the $\Lambda$CDM, $\sigma_8 = 0.8$  simulation.
The measured abundances from the three simulation box sizes 
64$h^{-1}$Mpc, 128$h^{-1}$Mpc and 256$h^{-1}$Mpc are the volume-weighted averages and errors over 8 realisations.
The volume conserving (\Vdn) model is shown as a black hatched (grey shaded) region
using $\deltav  = -2.7$ at $z=0.5$ ($z=1$) and the parameter range $\deltac = [1.06,1.686]$.
Note 
we have used only 
one colour for the results from the three simulation boxes at each redshift for clarity in this figure. 

We again find that the \Vdn\ model works very well in reproducing the abundance of voids in the dark matter in a $\Lambda$CDM 
universe at both redshifts, while the SVdW model, which is not plotted here for clarity, again 
overpredicts the abundances by approximately a factor of 5. 
Fig. \ref{fig:z_dep} shows that smaller (larger) voids are more (less) abundant at $z=1$ compared to $z=0.5$ which is also found in the model predictions
at both redshifts. 

\begin{figure}
\centering
{\epsfxsize=8.truecm
\epsfbox[ 80 360 436 685]{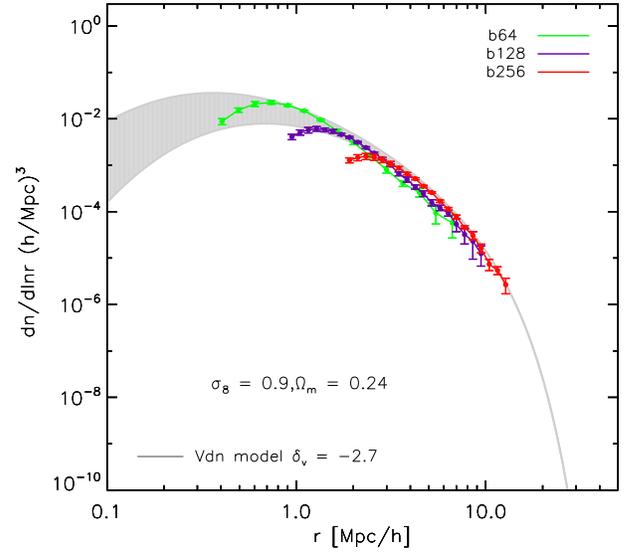}}
\vskip 0.25cm
{\epsfxsize=8.truecm
\epsfbox[ 80 360 436 685]{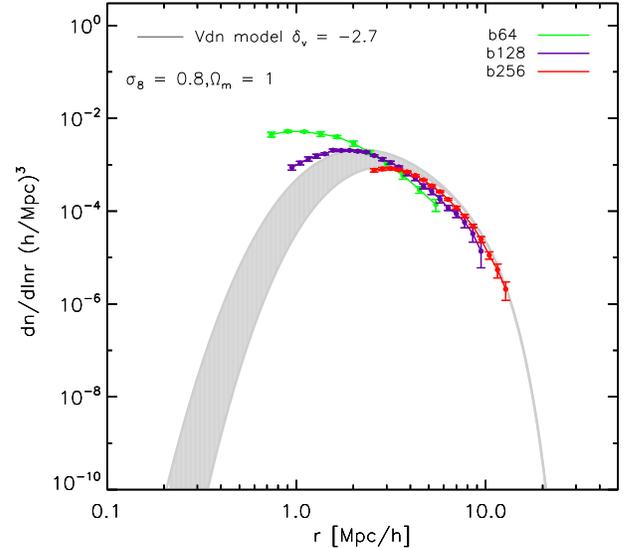}}
\caption{Void abundance for alternate cosmological parameters at $z=0$.  \emph{Upper:} $\Lambda$CDM with the initial conditions normalised to give  $\sigma_8 = 0.9$.
\emph{Lower:} EdS model with $\sigma_8 = 0.8$ with $\Omega_{\rm m}=1$.
The grey shaded region shows the \Vdn\ model within the parameter interval $\deltac$ = [1.06,1.686] and using $\deltav = -2.7$.
}
\label{s0.9}
\end{figure}

\subsection{Cosmological parameter variation \label{subsect:cosmology_dep}}

In order to check if the \Vdn\ model for the abundance of voids works when we change the cosmological model
we have run two simulations of alternative cosmologies to the standard $\Lambda$CDM with $\sigma_8 = 0.8$ which we discussed in the previous section.
In the first alternative cosmology we have chosen to modify only the value 
of $\sigma_8$ to $0.9$,  see Table \ref{table:simulations}; our second alternative 
cosmology is an Einstein  de-Sitter (EdS) universe where the matter density parameter $\Omega_{\rm m} = 1$. 
The linear perturbation theory power spectra for these simulations were generated using CAMB \citep{Lewis:2002ah} and 
normalised to $\sigma_8=0.9 $ ($\sigma_8=0.8$) for the $\Lambda$CDM (EdS) simulations in order to generate the initial conditions for the simulations and the variance
$\sigma(R)$ which is used in the excursion set model for the abundance.

The measured $z=0$ number density of voids with $\rhov=0.2 \rhom$ in 
the $\sigma_8=0.9$ and EdS simulations are shown in Fig. \ref{s0.9}.  
The volume conserving model is shown in both panels as a grey shaded region as in previous plots. 
We have used the same value, $\deltav = -2.7$, for the 
linear perturbation theory underdensity. 
Note this parameter is different in different cosmologies,
 however we find that such a small change in $\deltav$ going from an EdS to a $\Lambda$CDM universe
has a small impact on the predicted abundance of voids 
in the excursion set theory and the main differences arise from the change in the variance, $\sigma(R)$ (see Appendix \ref{sect:sphere}).

From Fig. \ref{s0.9},  it is clear that the volume conserving model works well in both of these cosmologies and fit the abundance of voids to within
25\% over the range $1<r(h^{-1}$Mpc$)<15$.
It is interesting to note the overall decrease in the 
abundance of voids in the dark matter distribution for voids with small radii $r<2 h^{-1}$Mpc in these two cosmologies 
which is most obvious in the measured number density from the EdS simulation
and a larger abundance for the $\sigma_8=0.9$ cosmology for large $r$.
 It is also clear from Fig. \ref{s0.9} (lower) that the excursion 
set model predicts more squashing of smaller voids due to the void-in-cloud effect but this
is occurring right on the 
resolution limit of our simulations at $2<r (h^{-1}$Mpc$)$.
Finally note that even if we modified the SVdW model in the ad hoc manner of Fig.~\ref{fig:nouniversal} to match the simulation results of $\Lambda$CDM with $\sigma_8=0.8$, the
predictions would be far off simulation results for the EdS cosmology.

\subsection{Halo defined voids \label{subsect:halo_voids}}

Voids in the galaxy population are defined not through the dark matter density field
but by the number density field $\nh$ of the dark matter halos they populate.
In this section
we  use  density minima in the halo number density. Our goal is to test how faithfully
 the abundance of voids in the dark matter matches that in the halo populations within the 
 context of the simple void finder of Section \ref{subsect:void_finder}.
It is important to note that a comparison between voids in the dark matter and halo 
distributions should account for the galaxy/dark matter halo biasing relation.
\citet{2003MNRAS.340..160B} showed that the properties of galaxy and dark matter voids differ significantly 
as a results of  galaxy bias e.g. if galaxies are sparse tracers of the underlying dark 
matter this give rise to larger voids in the galaxy distribution. \citet{2006MNRAS.366..467F} also studied the abundance of 
voids in the galaxy distribution within the excursion set formalism and showed that after accounting for bias, galaxy 
voids should be larger then dark matter voids, while voids selected 
using luminous galaxies should be larger then those using faint galaxies.

In this section we  use the FOF halo catalogues from the  128, 256, 500 $h^{-1}$Mpc simulation boxes and the publicly available
halo catalogues from the MultiDark and Bolshoi simulations \citep{2011arXiv1109.0003R}  which have computational 
box sizes of  $L=1000 h^{-1}$Mpc and $L=250 h^{-1}$Mpc on a side
respectively. These halos have been identified using the Bound-Density-Maxima algorithm \citep{1997astro.ph.12217K}.
We  only use halos which have $V_{\rm max}>200$km/s and $M > 1-2 \times 10^{12} h^{-1}M_\odot$ from the Bolshoi  and MultiDark simulations
to ensure that the statistics are robust.
We use the void finder described in Section \ref{subsect:void_finder} to identify voids in the distribution of halos
which have $\nv = 0.2\nh$ where $\nv$ is the average number density in the void whereas $\nh$ is the average in the whole simulation.
Our final sample consists of 5,768 voids  using 1.7 $\times 10^{6}$ halos from the MultiDark simulation
and 4,826 voids using 2.2 $\times 10^6$ halos from the Bolshoi simulation.
Both of these
 simulations are of a higher resolution than the ones we carried out in 128, 256, 500 $h^{-1}$Mpc simulation boxes -- it is useful to compare the abundance
of voids in the halo population from these simulations to ours as an indication of the scales at which our results have converged.

\begin{figure}
\centering
{\epsfxsize=8.truecm
\epsfbox[ 80 360 436 685]{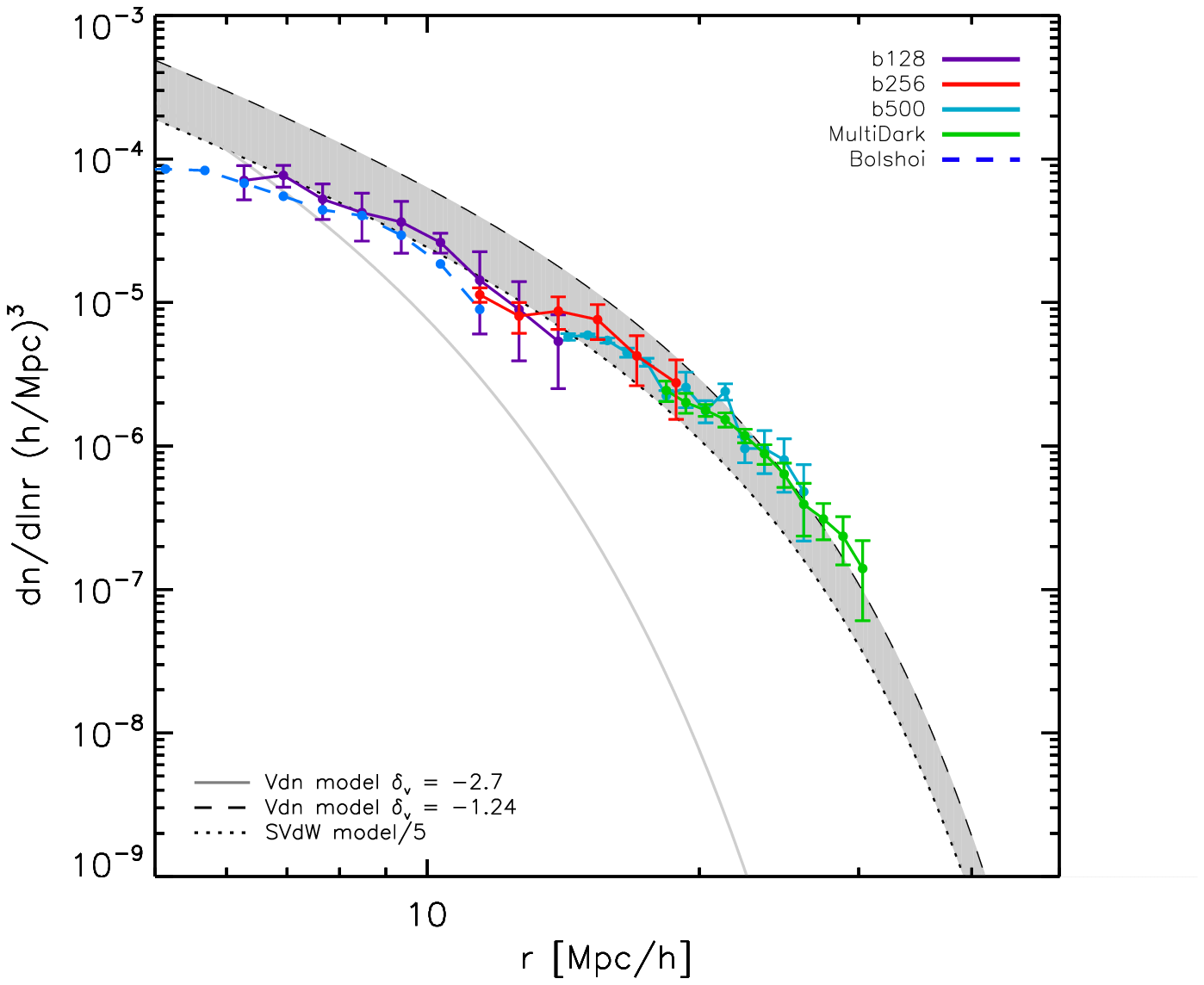}}
\caption{The number density of voids with  $\nv = 0.2 \nh$ in the halo distribution
from the  128 (purple), 256 (red), 500 (cyan) $h^{-1}$Mpc simulation boxes. The results from the MultiDark (Bolshoi) simulation are shown in dark green (blue).
The error bars represent the error on the mean from eight simulations. The errors on the MultiDark simulation represent the Jackknife error on the mean.
The grey shaded region bounded by the black dashed and dotted line represents the volume conserving model with $\deltav = -1.24$ and varying amplitude
as in Fig. \ref{rho_0.3}. The grey solid line represents the \Vdn\ model with $\deltav = -2.7$.
}
\label{fig:halos}
\end{figure}

The measured abundance of voids in the halos population from our simulations and the Bolshoi and MultiDark catalogues at $z=0$ are shown in Fig. \ref{fig:halos}.
The errors shown  on the results from the 128 (purple), 256 (red), 500 (cyan) $h^{-1}$Mpc simulation boxes are measured from the scatter amongst eight
different realisations in each box size. The errors on the MultiDark  simulation  were obtained by jackknife sampling from
each  simulation by dividing the simulation volume
into $N_{\rm sub}= 8$ equal subvolumes and then systematically
omitting one subvolume at a time in order to calculate
the void abundance on the remaining $N_{\rm sub}-1$ volume.
We find that voids identified in this manner through the halo distribution do not follow
the \Vdn\ model assuming $\nv = 0.2\nh$ which corresponds to dark matter voids
of $\rhov=0.2\rhom$.  They also do not follow the SVdW model which would have the same
shape but 5 times the amplitude.

\begin{figure*}
\centering
{\epsfxsize=13.truecm
\epsfbox[ 77 380 537 632]{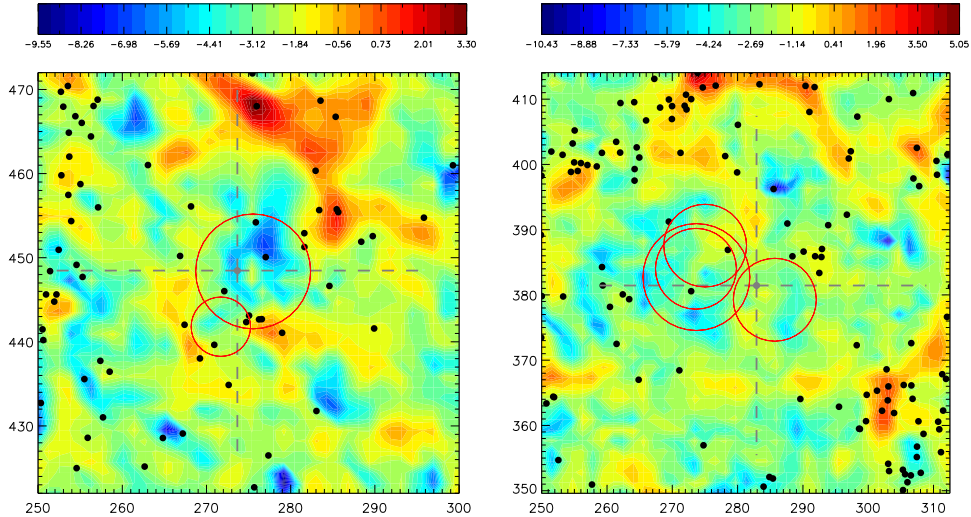}}
\caption{ Left: A $10\times50\times50h^{-1}$Mpc slice through the
 500 $h^{-1}$Mpc simulation box centred on a large, $r\sim 21 h^{-1}$Mpc, void in the halo population (black dots). The diameter of the void is shown as a dashed grey line
and the coloured contours represent the log of the densify field which has been evaluated on a grid of $256^3$ points.
The red circles represent all the voids in dark matter which have $\rhov = 0.2 \rhom$
and whose centres are within 10$h^{-1}$Mpc of the centre of the void in the halo distribution.
Right: A $60\times14\times60h^{-1}$Mpc slice through the
 500 $h^{-1}$Mpc simulation box centred on a large, $r\sim 26 h^{-1}$Mpc, void in the halo population (black dots).
The red circles represent all the voids in dark matter which have $\rhov = 0.2 \rhom$ and whose centres are within 10$h^{-1}$Mpc of the centre of the void in the halo distribution. Note these voids only appear to be overlapping due to the projection effect.
}
\label{fig:contour}
\end{figure*}

Overall Fig. \ref{fig:halos} shows that our void finder finds large 
halo defined voids that do not correspond to dark matter defined voids of the same underdensity for $r \ga 10h^{-1}$Mpc. Although it is difficult to compare these results with previous work due to the large differences in the void finders used,
qualitatively this agrees with the findings of \citet{2003MNRAS.340..160B} who
 measured the void probability function from simulations and found that the VPF for voids with $r>5 h^{-1}$Mpc was much higher for the galaxy catalogues compared to the dark matter.
These results illustrate the fact that there is not always
a 1:1 correspondence between voids in the dark matter and the dark matter
halo distributions and this is especially pronounced when we define a void as having a fixed underdensity which is the same for dark matter and halos.

To illustrate the mismatch between the voids which we find in the dark matter and halo distributions using the same underdensity criterion, Fig. \ref{fig:contour}
shows a $10\times50\times50h^{-1}$Mpc (left panel) and a $60\times14\times60h^{-1}$Mpc (right panel) slice through the
dark matter density field which has been evaluated on a grid
of $256^3$ points from
 the 500 $h^{-1}$Mpc simulation box. The coloured contours represent the log of the
density field in each cell and the halos around each void are represented by black dots.
The radius of each void is $r\sim 21 h^{-1}$Mpc (left panel) and $r\sim 26 h^{-1}$Mpc (right panel) and is shown as a grey dashed line in Fig. \ref{fig:contour}.
The red circles in this plot show the voids identified in the dark matter whose centres are within $10 h^{-1}$Mpc of the centre of the void in the halo population.
Not only 
is it possible to find more than one dark matter 
void which overlaps with the halo void but the radii of the dark matter voids at which $\rhov = 0.2 {\rhom}$
are a lot smaller than the halo voids which satisfy the analogous criterion.

There are at least two possible ways to reconcile the \Vdn\  predictions for the abundance of dark matter voids with that of the halo voids.
Firstly, a scale dependent modification to the barrier in the \Vdn\ model could be used to alter the underdensity threshold used to find voids in the dark matter.
Secondly if we keep a fixed underdensity threshold  to define dark matter voids, it may be possible to find a scaling of this threshold
to define voids in the halo distribution.
These ideas are beyond the scope of this work but see
\citet{2006MNRAS.366..467F} for related ideas.

As a simpler illustration of these ideas, 
 in Fig.  \ref{fig:halos} we also plot the \Vdn\ model assuming that halo defined voids of $\nv = 0.2\nh$ correspond to dark matter
defined voids of 
 $\rhov = 0.4\rhom$. 
These predictions are plotted as a grey shaded region
allowing the amplitude to vary from the predictions of the \Vdn\ model which
rescales the SVdW amplitude by $\rhov/\rhom=0.4$ (black dashed lines) and the rescaling of 1/5 that fits our dark matter voids well  as in Fig. \ref{rho_0.3}
(black dotted line).
Compared to the predictions of the \Vdn\ model for dark matter voids of $\nv = 0.2\nh$ (solid grey line), these black dashed and dotted curves
match the abundance of voids in the halo populations better though no single rescaling
matches perfectly across the full range.

\section{Summary and Conclusions}
 \label{sect:conclusions}

The next generation of galaxy redshift surveys such as BigBOSS \citep{2009arXiv0904.0468S}, 
Euclid \citep{2011arXiv1110.3193L} and {\sc WFIRST} \citep{2006astro.ph..9591A,2012arXiv1208.4012G}
will allow us to study the large scale structure of our Universe in ever greater detail.
Cosmic voids represent one of the main components which strongly influence the growth
of clusters, walls and filaments in the mass distribution.
Studying the statistics and dynamics of these underdense regions is a promising 
way to test the cosmological model and
 hierarchical structure formation.

Several challenges which may affect the usefulness of voids as a probe of cosmology are addressed in this paper, such as, 
given a survey or numerical simulation of a given size,
 how robust are the statistics on the number density of voids of a given size, how accurately can we predict 
the number density of these voids and 
how faithfully do voids in the halo population represent voids in the dark matter.
Using N-body simulations of a $\Lambda$CDM cosmology we test the excursion set model for the abundance of voids including the 
model provided by \citet{2004MNRAS.350..517S}, which takes into account the void-in-void and void-in-cloud scenarios. Our void finder
makes use of the {\sc ZOBOV} \citep{2008MNRAS.386.2101N} algorithm which uses Voronoi tessellation to locate density minima.
We define a void as a spherical region around these minima with $\rhov = 0.2 \rhom$ and make use of several different computational 
box sizes so we can 
determine the volume and resolution that is needed in order to recover robust statistics for voids of a given size.
We have tested the robustness of our void finder to the following changes and have found convergent results:
using different simulation boxsizes and particle numbers, using a volume-weighted centre or the core particle to define the centre of a void,
using all particles around the density minima or only particles in a zone given by {\sc ZOBOV} and using only statistically significant 
voids or voids with a core particle density $<0.2 \rhom$.

We find that the measured abundance of voids at 
$z=0$ from a $\Lambda$CDM simulation 
does not match the predictions of the \citet{2004MNRAS.350..517S} model, which greatly 
overpredicts the results using a linear underdensity of
$\deltav = -2.7$. 
We find that the excursion set theory, which is the basis of the SVdW model, accurately predicts the shape of the abundance of voids measured from simulations. However the predicted amplitude is incorrect due to the assumption of isolated spherical expansion which does not account for the merging of voids as they expand.
Instead we find a volume conserving model, which is also based on the excursion set method with
$\deltav = -2.7$, matches the measured abundances to within 16\% for void radii $1< r(h^{-1}$Mpc$)<15$.
This model works remarkably well and suggests that the number
density of voids decreases in going from the linear to the nonlinear regime by the same amount that the voids expand.   This agreement is robust to varying the
redshift in the $\Lambda$CDM model as well as the underlying cosmology.  
Using simulations of different cosmological models, a $\Lambda$CDM cosmology
 with $\sigma_8=0.9$ and a Einstein-deSitter cosmology with  $\Omega_{\rm m} =1$, we find that the volume conserving model works well and reproduces the measured number density
from each simulation to within 25\% over the range $1< r(h^{-1}$Mpc$)<15$.
We have also tested model predictions for 
density thresholds of $\rhov = 0.3\rhom$
and $\rhov = 0.4\rhom$ and find that the volume fraction physicality rescaling factor
remains fixed at $\sim 1/5$ rather than scaling as $\rhov/\rhom$.

Using the number density threshold criteria of $\nv = 0.2\nh$ in our void finder we have examined the voids in the halo population from the 128, 256 
and the 500$h^{-1}$Mpc on a side computational boxes. We also use the Bolshoi and the Multidark \citep{2011arXiv1109.0003R} simulations to measure void abundances. 
These two simulations are of a  higher resolution than our simulations and we have confirmed that our measured void abundances in the halo population agree with both the 
MultiDark and Bolshoi measurements.
We find that the number density of voids in the dark matter 
halo distribution is very different to that in the dark matter, with fewer small, $r<10 h^{-1}$Mpc, voids and many more large, $r>10 h^{-1}$Mpc,
voids.
These results indicate that a
 given void in the halo distribution of fixed underdensity of $\nv = 0.2\nh$
cannot be unambiguously related to a void in the dark matter of equal underdensity and in the case that there is a 1:1 correspondence, the radii at which a dark matter
or halo void have a given underdensity can be very different.

Cosmic voids are a promising and interesting tool which can be used to test many aspects of the $\Lambda$CDM cosmological model and having an accurate
model for the number density of voids in the Universe represents a first step.
In this work we have presented a model based on the excursion set theory which conserves the volume fraction of voids and works well 
at predicting 
the abundance of voids identified in dark matter from N-body simulations over a range of scales. Establishing the relationship between voids in the dark matter and in the halo or
galaxy distribution requires further study.
\section*{Acknowledgments}

EJ acknowledges the support of a grant from the Simons Foundation, award number 184549. This work was supported in part by the Kavli Institute for Cosmological Physics at the University of Chicago through grants NSF PHY-0114422 and NSF PHY-0551142 and an endowment from the Kavli Foundation and its founder Fred Kavli.
YL and WH were additionally supported by U.S.~Dept.\ of Energy contract DE-FG02-90ER-40560 and the David and Lucile Packard Foundation.
We are grateful for the support of the University of Chicago Research Computing Center for assistance with the calculations carried out in this work.

\appendix

\section{Spherical evolution model} \label{sect:sphere}

In this Appendix we review the spherical evolution model
which describes the nonlinear evolution of an un-compensated spherically
symmetric tophat underdense (overdense) perturbation.
To illustrate the physics, we will use the spherical tophat model
in Einstein-de Sitter cosmology as an example,
which is analytically solvable until shell crossing.

Consider an initial spherical tophat density perturbation ($|\delta_0|\ll 1$) of physical radius $R_0$
at $a_\text{i}=a(t_\text{i})$.
We can think of the perturbation as composed of concentric mass shells,
labeled by their initial physical radii $R_\text{i}$ at $a_\text{i}$.
Let $\Delta(R_\text{i},a)$ denote the average overdensity of the region enclosed by
the mass shell $R_\text{i}$. Then its initial value is
\ba
    \Delta_\text{i}(R_\text{i}) \equiv \Delta(R_\text{i}, a_\text{i}) = \left\{ \begin{array}{ll}
        \delta_0 & \quad R_\text{i} \leq R_0, \\
        \delta_0 (R_0/R_\text{i})^3 & \quad R_\text{i} > R_0.
    \end{array} \right.
\ea
For brevity we omit the $R_\text{i}$ argument of $\Delta_\text{i}$ below.
According to Birkhoff's theorem, the evolution of the mass shell $R_\text{i}$ only depends on the total mass 
inside $R_\text{i}$, but not the mass distribution or the mass outside.
Thus the shell $R_\text{i}$ evolves in the same way as a FLRW universe
\ba
\label{eqn:EdS}
    \left[\frac{\dot R(t;R_\text{i})}{R(t;R_\text{i})}\right]^2
    = H_\text{i}^2 \left[ ( 1 + \Delta_\text{i} ) \left( \frac{R_\text{i}}{R} \right)^{3}
    - \frac53 \Delta_\text{i} \left( \frac{R_\text{i}}{R} \right)^{2} \right],
\ea
where $R(t)$ is the physical radius, and initial conditions are set to the growing mode in linear theory.
By introducing the dimensionless conformal time
\ba
    \text{d}\eta = \frac{R_\text{i}}{R}
    \sqrt{ \left| \frac53 \Delta_\text{i} \right| } \, H_\text{i} \text{d}t \, ,
\ea
we can solve equation \eqref{eqn:EdS} in a parametric form (to  leading order in $\Delta_\text{i}$)
\ba
    \frac{R}{R_\text{i}} & \simeq & \frac12 \left|\frac53 \Delta_\text{i} \right|^{-1}
    \begin{cases}
         (\cosh\eta - 1) &\quad \delta_0<0, \\ 
         (1 - \cos\eta) &\quad \delta_0>0; \\
    \end{cases}  \label{eqn:R} \\
    H_\text{i} t & \simeq & \frac12 \left| \frac53 \Delta_\text{i} \right|^{-\frac32}
   \begin{cases}
         (\sinh\eta - \eta) &\quad \delta_0<0, \\ 
         (\eta - \sin\eta) &\quad \delta_0>0.\\
    \end{cases} \label{eqn:t}
\ea

Given this evolution of mass shells, we are particularly interested
in  shell crossing for the expansion case ($\delta_0<0$),
which is usually seen as a characteristic event that signifies
the formation of the void at a nonlinear level.
Note that these solutions represent a family of trajectories labeled by $R_\text{i}$ and parametrized by $\eta_{R_{\text{i}}}$.
We can find out when and where shell crossing first occurs by differentiating the parametrized solutions 
with respect to $R_\text{i}$ and $\eta$,
and requiring that $\text{d}R$ and $\text{d}t$ vanish, for $R_\text{i}>R_0$
\ba \label{eqn:sc}
    \begin{bmatrix}
        A_{11} & A_{12} \\
        A_{21} & A_{22}
    \end{bmatrix}
    \begin{bmatrix}
        \text{d}R_\text{i}/R_\text{i} \\
        \text{d}\eta
    \end{bmatrix}
    = 0 ,
\ea
where
\ba
    A_{11} &=& 2 \left| \frac53 \Delta_\text{i} \right|^{-1} (\cosh\eta - 1) ,\nonumber \\
    A_{12} &=& \frac12 \left| \frac53 \Delta_\text{i} \right|^{-1} \sinh\eta , \nonumber\\
    A_{21} &=& \frac94 \left| \frac53 \Delta_\text{i} \right|^{-\frac32} (\sinh\eta - \eta) , \nonumber\\
    A_{22} &=& \frac12 \left| \frac53 \Delta_\text{i} \right|^{-\frac32} (\cosh\eta - 1) .
\ea
For this homogeneous system of linear equations to have nonzero solutions,
we must have $\det A=0$. Thus we derive the shell crossing condition
\ba \label{eqn:89}
    \frac{\sinh\eta \, (\sinh\eta-\eta)}{(\cosh\eta-1)^2} = \frac89 .
\ea
Shell crossing first happens at $\eta_\text{sc} = 3.488$ among the boundary shells, 
i.e., $R_\text{i}=R_0$ in the above criterion. At shell crossing, the void interior
has a relative density
\ba
    1+\Delta_\text{sc} \simeq \frac{9}{2}
    \frac{(\sinh\eta_\text{sc}-\eta_\text{sc})^2}{(\cosh\eta_\text{sc}-1)^3} = 0.2047 \, ,
\ea
which implies that the void has expanded by a factor of $(1+\Delta_\text{sc})^{-1/3} = 1.697$
in comoving radius. Note that these numbers do not depend on the size of the void.

To calculate the linear theory prediction of the void underdensity at shell crossing $\delta_\text{v}$,
we expand $R(t)$ to the first order with the help of the parametric solution \eqref{eqn:R} and \eqref{eqn:t},
for $R_\text{i}\leq R_0$
\ba
    \frac{R}{R_\text{i}} = \frac{a}{a_\text{i}}
    \left[1 - \frac{\delta_0}{3} \left(\frac32 H_\text{i} t\right)^\frac23
    + \cdots \right] ,
\ea
the first order of which gives the linear underdensity
\ba
    \delta = \delta_0 \left(\frac32 H_\text{i} t\right)^\frac23
    \simeq - \frac{3}{20} \big[6(\sinh\eta-\eta)\big]^{2/3} .
\ea
Thus the linear underdensity at shell crossing is
\ba
    \delta_\text{v} = \delta(\eta_\text{sc}) = -2.717 \, .
\ea
Note this number is different in different cosmologies, e.g it is
$\delta_\text{v} = -2.731$ for $\Lambda$CDM. However we find that such a small
change in $\delta_\text{v}$ going from an EdS to a $\Lambda$CDM universe has a
small impact on the predicted abundance of voids in the excursion set theory.
Also note the value for $\delta_\text{v}$ for the EdS universe 
quoted here is a correction to the result presented in
\citet{2004MNRAS.350..517S} ($\delta_{\rm v} = -2.81$).

Similarly, for the spherical collapse model
\ba
    \delta = \delta_0 \left(\frac32 H_\text{i} t\right)^\frac23
    \simeq \frac{3}{20} \big[6(\eta-\sin\eta)\big]^{2/3} .
\ea
The well-known turn-around and virialization of halos occur at
$\eta_\text{ta}=\pi$ and $\eta_\text{vir}=2\pi$,
leading to $\delta_\text{c}=1.062$ and $\delta_\text{c}=1.686$ respectively.
For the $\Lambda$CDM model these become $\delta_\text{c}=1.303$ and
$\delta_\text{c}=1.674$, for turnaround and collapse at $z=0$.
The EdS range encompasses that of the
$\Lambda$CDM parameters and so in the main text we have adopted the EdS parameters
to show the full range of possibilities.

Finally note that these linear density thresholds $\delta_\text{v}$ and $\delta_\text{c}$,
which are to be used in the excursion set formalism, are independent of the size of the structures.

\begin{figure}
\centering
{\epsfxsize=8.truecm
\epsfbox{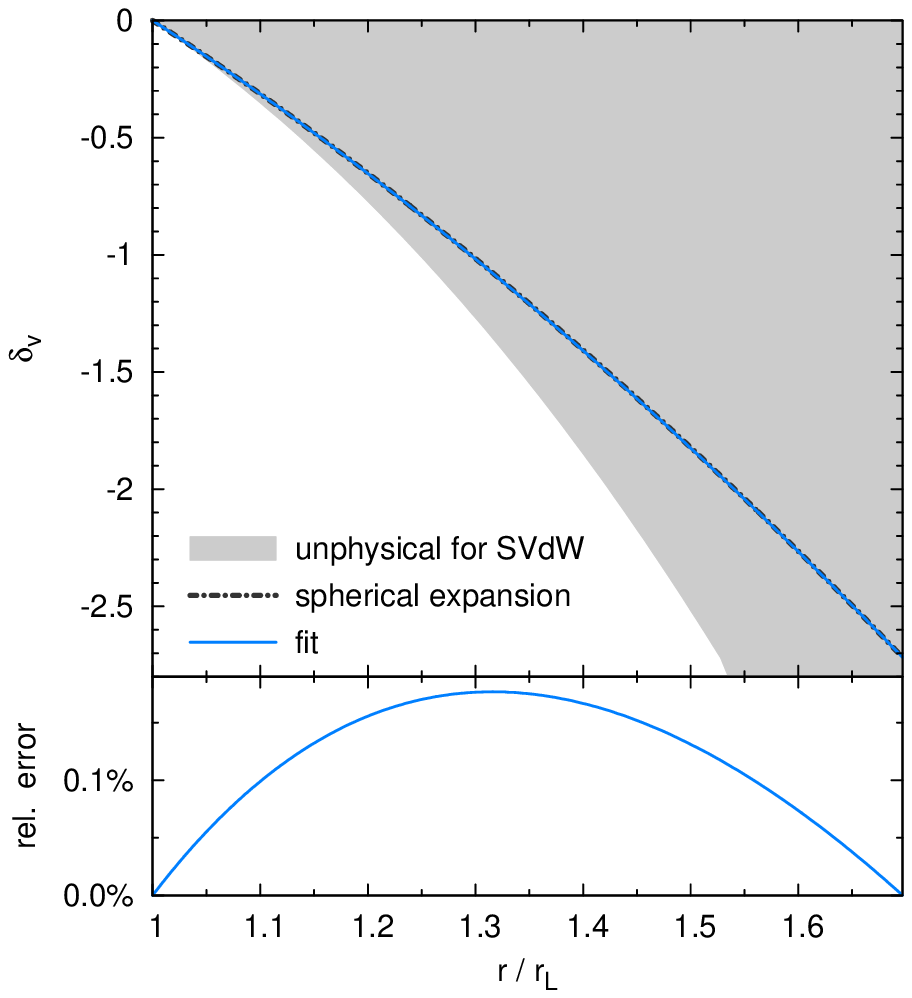}}
\caption{Linear underdensity as a function of  void expansion factor  (exact: dot-dashed line,
fit of equation \eqref{eqn:deltavrNL}: 
blue solid line).  
Also shown is the physicality constraint from requiring that the total volume
fraction in voids from equation\ \eqref{eqn:totvol} remain less than unity in SVdW model for
the least stringent value of $\deltac=1.06$ (shaded region).   Only as $|\deltav| \rightarrow 0$, where the
regions are at the mean density and do not expand, does the
model regain physicality.  Bottom panel shows fractional error in the fit.
}
\label{fig:alwaysfillup}
\end{figure}

\section{SVdW model modifications} \label{sect:mod}

In this Appendix we explore modifications of the SVdW model prescription and
approximations introduced in \cite{2004MNRAS.350..517S}.
The spherical evolution model relates the linear underdensity $\delta$ to
nonlinear underdensity $\Delta$, or alternatively to
$\rNL/\rL=(1+\Delta)^{-\frac13}$, where $\rNL$ is the void radius $\rNL=R(t;R_0)/a(t)$
and $\rL$ the linear radius $\rL=R_0/a_\text{i}$. 
If we relax the criterion for defining a void to correspond to underdense regions that
have undergone shell crossing, there is additional freedom in defining the
void abundance as a function of radius so long as $\deltav$ and $\rNL/\rL$ are
chosen self-consistently.

We show this relation  before shell crossing for the EdS model
in Fig. \ref{fig:alwaysfillup}.   The relation is well fit to \citep{1994ApJ...427...51B}
\ba
\deltav \approx c [1 - (\rNL/\rL)^{3/c}] = c [1 - (\rhov/\rhom)^{-1/c}],
\label{eqn:deltavrNL}
\ea
where $c = 1.594$, with errors below $0.2\%$.
Also shown is the maximum $r/\rL$, constrained
by requiring the total volume fraction in voids from equation\ \eqref{eqn:totvol} be less than
$1$, in the SVdW model. Clearly this constraint depends on $\delta_\text{c}$, which
in the plot is chosen to be the least stringent value $1.06$ in the expected
range. Note that no choice of $\delta_\text{v}$ and $\rNL/\rL$ is physical
for they all violate the total volume condition.
In the main text, we also considered ad hoc modifications of the model where
$\deltav$ and $\rNL/\rL$ are considered unrelated.

\begin{figure}
\centering
{\epsfxsize=8.truecm
\epsfbox{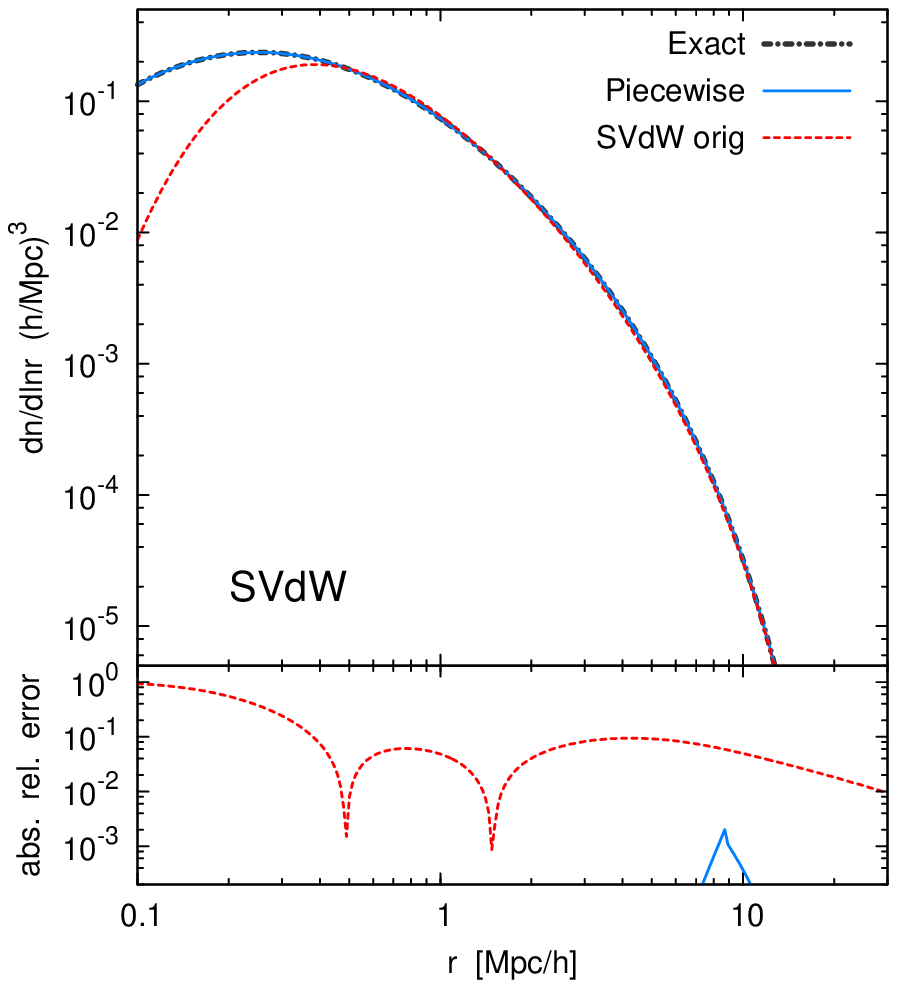}}
\caption{SVdW model void abundance is shown with three different
$f_{\ln\sigma}$: the exact formula  \eqref{eqn:exactf} 
(thick black dot-dashed
line), the \citet{2004MNRAS.350..517S} SVdW approximation
\eqref{eqn:approxforig} (red dashed line), and our piecewise approximation
\eqref{eqn:approxf} (blue solid line), along with the absolute value of the error relative to exact. We use $\delta_\text{v}=-2.7$ and
$\delta_\text{c}=1.06$, well within the stated domain of validity of both approximations.
The SVdW approximation breaks down where the void-in-cloud process
becomes important, whereas the piecewise approximation is accurate everywhere
with errors peaking  at $0.2\%$ where the piecewise transition occurs.
We use the $\sigma_8=0.8$ $\Lambda$CDM
cosmology as listed in Table \ref{table:simulations}.
}
\label{fig:svdwapprox}
\end{figure}

\cite{2004MNRAS.350..517S} also utilised an approximation to the exact prediction
for the abundance function of equation \eqref{eqn:exactf} which introduces notable errors for
scales where the void-in-cloud process dominates and consequently the total volume
fraction.   Their approximation
\ba
\label{eqn:approxforig}
     f_{\ln \sigma}(\sigma)   &\approx& \sqrt{\frac{2\nu}{\pi}}\exp\left(-\frac{\nu}{2}\right) \exp\left( -\frac{|\delta_{\rm v}|}{\deltac}\frac{\mathcal{D}^2}{4\nu} - 2\frac{\mathcal{D}^4}{\nu^2}\right) ,
\ea
where 
$\nu  = \delta_\text{v}^2/\sigma^2(M)$,
had a stated realm of validity of $\delta_\text{c}/|\delta_\text{v}|>1/4$ or $\mathcal{D}<4/5$.
Unfortunately, this approximation has uncontrolled errors at $\nu \ll 1$, exactly where
the void-in-cloud process operates as shown in Fig.~\ref{fig:svdwapprox}. 
Our piecewise approximation is accurate at the $0.2\%$ level or better everywhere.
Note that the errors and smoothness of our approximation can be improved at the transition
point by a suitable interpolation between the two piecewise curves though it is not
necessary for this work.

\bibliographystyle{mn2e_hyperref}
\bibliography{thebibliography}

\bsp

\label{lastpage}

\end{document}